\documentclass[usenatbib,useAMS,usedcolumn]{mn2e}

\usepackage{epsfig}
\usepackage{graphics,graphicx}
\usepackage{times}
\pdfoutput=1

\newcommand{\arcm}{\hbox{$^\prime$}}

\newcommand{\degree}{\hbox{$^\circ$}}
\newcommand{\rosat}{\emph{ROSAT}}
\newcommand{\chandra}{\emph{Chandra}}
\newcommand{\xmm}{\emph{XMM-Newton}}
\newcommand{\xmms}{\emph{XMM}}
\newcommand{\asca}{\emph{ASCA}}

\newcommand{\gmrt}{\emph{GMRT}}
\newcommand{\vla}{\emph{VLA}}
\newcommand{\arcs}{\mbox{\arcm\arcm}}

\newcommand{\Zsol}{\ensuremath{Z_{\odot}}}
\newcommand{\Lsol}{\ensuremath{L_{\odot}}}
\newcommand{\Msol}{\ensuremath{~M_{\odot}}}

\newcommand{\s}{\ensuremath{\mbox{~s}}}
\newcommand{\ps}{\ensuremath{\s^{-1}}}
\newcommand{\yr}{\ensuremath{\mbox{~yr}}}
\newcommand{\pyr}{\ensuremath{\yr^{-1}}}
\newcommand{\Msolpyr}{\ensuremath{\Msol \pyr}}
\newcommand{\cm}{\ensuremath{\mbox{~cm}}}
\newcommand{\pcmsq}{\ensuremath{\cm^{-2}}}
\newcommand{\pcmcu}{\ensuremath{\cm^{-3}}}
\newcommand{\km}{\ensuremath{\mbox{~km}}}

\newcommand{\erg}{\ensuremath{\mbox{~erg}}}
\newcommand{\ergps}{\ensuremath{\erg \ps}}
\newcommand{\ergpspcmsq}{\ensuremath{\erg \ps \pcmsq}}
\newcommand{\kmps}{\ensuremath{\km \ps}}

\newcommand{\ltsim}{\,\rlap{\raise 0.5ex\hbox{$<$}}{\lower 1.0ex\hbox{$\sim$}}\,} 
\newcommand{\gtsim}{\,\rlap{\raise 0.5ex\hbox{$>$}}{\lower 1.0ex\hbox{$\sim$}}\,} 

\newcommand{\Hi}{H\textsc{i}}

\begin{document}

\title[ 
A type 2 QSO in a cooling flow
] 
{ 
A Giant Metrewave Radio Telescope/Chandra view of IRAS~09104+4109: A type 2 QSO in a cooling flow
}

\author[E. O'Sullivan et al.]  {Ewan O'Sullivan\footnotemark[1]$^{1,2}$,
  Simona Giacintucci$^{3,4}$, Arif Babul$^{5}$, Somak Raychaudhury$^{1}$,
  \newauthor Tiziana Venturi$^{6}$, Chris Bildfell$^{5}$, Andisheh
  Mahdavi$^{7}$, J.~B.~R. Oonk$^{8}$, Norman Murray$^{9}$
  \newauthor Henk Hoekstra$^{10}$ and Megan Donahue$^{11}$\\
  $^{1}$ Harvard-Smithsonian Center for Astrophysics, 60 Garden Street,
  Cambridge, MA 02138, USA \\
  $^{2}$ School of Physics and Astronomy, University of Birmingham,
  Edgbaston, B15 2TT, UK \\
  $^{3}$ Department of Astronomy, University of Maryland, College Park, MD
  20742-2421, USA \\
  $^{4}$ Joint Space-Science Institute, University of Maryland, College
  Park, MD 20742-2421, USA \\
  $^{5}$ Department of Physics and Astronomy, University of Victoria,
  Victoria, BC V8P 1A1, Canada\\
  $^{6}$ INAF, Istituto di Radioastronomia, Via Gobetti 101, 40129 Bologna,
  Italy\\
  $^{7}$ Department of Physics and Astronomy, San Francisco State
 University, San Francisco, CA 94131, USA\\
  $^{8}$ Netherlands Institute for Radio Astronomy (ASTRON), Postbus 2, 7990 AA Dwingeloo, The Netherlands\\
  $^{9}$ Canadian Institute for Theoretical Astrophysics, 60 St. George Street, University of Toronto, ON M5S 3H8, Canada\\
  $^{10}$ Leiden Observatory, Leiden University, P.O. Box 9513, NL-2300 RA
 Leiden, The Netherlands\\
  $^{11}$ Physics and Astronomy Department, Michigan State University, East Lansing, MI 48824-2320, USA}

\date{Accepted 2012 June 2; Received 2012 May 31; in original form 2012 April 9}

\pagerange{\pageref{firstpage}--\pageref{lastpage}} \pubyear{2012}

\maketitle

\label{firstpage}

\begin{abstract} 
  IRAS~09104+4109 is a rare example of a dust enshrouded type 2 QSO in the
  centre of a cool--core galaxy cluster. Previous observations of this
  $z$=0.44 system showed that as well as powering the hyper--luminous
  infrared emission of the cluster--central galaxy, the QSO is associated
  with a double--lobed radio source. However, the steep radio spectral
  index and misalignment between the jets and ionised optical emission
  suggested that the orientation of the QSO had recently changed.  We use a
  combination of new, multi-band \textit{Giant Metrewave Radio Telescope}
  observations and archival radio data to confirm that the jets are no
  longer powered by the QSO, and estimate their age to be 120-160~Myr. This
  is in agreement with the $\sim$70-200~Myr age previously estimated for
  star--formation in the galaxy. Previously unpublished \textit{Very Long
    Baseline Array} data reveal a 200~pc scale double radio source in the
  galaxy core which is more closely aligned with the current QSO axis and
  may represent a more recent period of jet activity.  These results
  suggest that the realignment of the QSO, the cessation of jet activity,
  and the onset of rapid star--formation may have been caused by a
  gas--rich galaxy merger.  X--ray observations reveal a spiral structure
  in the ICM which suggests that the cluster is in the process of
  relaxation after a tidal encounter or merger with another system; such a
  merger could provide a mechanism for transporting a gas--rich galaxy into
  the cluster core without stripping its cold gas.  A \chandra\ X--ray
  observation confirms the presence of cavities associated with the radio
  jets, and we estimate the energy required to inflate them to be
  $\sim$7.7$\times$10$^{60}$~erg. The mechanical power of the jets is
  sufficient to balance radiative cooling in the cluster, provided they are
  efficiently coupled to the intra--cluster medium (ICM).  We find no
  evidence of direct radiative heating and conclude that the QSO either
  lacks the radiative luminosity to heat the ICM, or that it requires
  longer than 100-200~Myr to significantly impact its environment.

\end{abstract}

\begin{keywords}
galaxies: clusters: individual (CL~09104+4109) --- galaxies: individual
(IRAS~09104+4109) --- galaxies: quasars: general --- X--rays:
galaxies: clusters --- galaxies: active --- cooling flows
\end{keywords}

\footnotetext[1]{E-mail: ejos@star.sr.bham.ac.uk}

\section{Introduction}
\label{sec:intro}

It is now widely agreed that feedback from active galactic nuclei (AGN) is
likely to be the dominant process governing star formation and gas cooling
in systems ranging from large galaxies to the most massive galaxy clusters
\citep{McNamaraNulsen07,PetersonFabian06}. In the nearby Universe there is
a wealth of evidence that radio galaxies in groups and clusters have an
impact on the surrounding intra-cluster medium (ICM) by inflating cavities
\citep[e.g.,][]{Raffertyetal06,Birzanetal08,Dunnetal10,OSullivanetal11b},
driving shocks and sound waves
\citep[e.g.,][]{Nulsenetal05,Fabianetal06,SandersFabian07,Formanetal07},
and causing gas mixing
\citep[e.g.,][]{Fabianetal05,Wiseetal07,Blantonetal09,Simionescuetal09,Kirkpatricketal11}.

While a detailed, universally accepted description of the mechanisms
underlying the origin and power of AGN jets remains elusive, the
combination of the results from numerous analytic studies and increasingly
sophisticated numerical simulation studies \citep[e.g.,][]{ Koide03,
  Gammieetal04, DeVillersetal05,HawleyKrolik06, Komissarov07, Punsly07,
  Beckwithetal08, McKinneyBlandford09} do suggest an emerging consensus on
a few key elements of the picture \citep[c.f.][for a more detailed
overview]{BensonBabul09}:

First, jets are an ubiquitous feature of magnetized accretion flows onto
black holes \citep{DeVillersetal05,Punsly07,McKinneyBlandford09}. The jets
may be highly relativistic, narrowly collimated and Poynting-flux dominated
or broad, mildly relativistic and matter-dominated.

Second, the launching of powerful jets is most efficient when the accretion
flow in the vicinity of the black hole has a large scale-height, as is the
case when the flow is radiatively inefficient and advection-dominated
\citep[e.g.,][]{Meier01,Churazovetal05,Nemmenetal07}.  Radiatively
inefficient flows are believed to arise at low accretion rates (i.e
$\dot{M} < 0.01 \dot{M}_{Edd}$) \citep[e.g.,][]{NarayanYi94}.  Above this
threshold accretion rate, the flow transitions to a radiatively efficient
thin disk.  There is a considerable body of work suggesting that the jet
production is suppressed in thin disks
\citep{Livioetal99,Meier01,Maccaroneetal03}.  At very high accretion rates
(i.e $\dot{M} > 0.1-0.5 \dot{M}_{Edd}$), the nature of the flow and even
its radiative properties are not well understood, though there are
suggestions that as the flow becomes optically thick, the disk scale height
increases and powerful jets may be launched.

Third, the outflows draw their energy from both the gravitational energy
released by the accretion flow as well as the rotational energy of the
black hole itself.  This link between black hole spin and outflow power has
long been indicated from theoretical considerations
\citep[c.f.,][]{BlandfordZnajek77,PunslyCoronitti90,Meier99,Meier01} and in
recent years, has received further support from numerical studies
\citep[c.f.,][]{HawleyKrolik06,Komissarov07, Punsly07,McKinneyBlandford09}.
MHD simulations show that the magnetic fields anchored in the rotating
accretion flow will give rise to both an outflow of material from the
accretion flow and the surrounding corona, as well as a hydromagnetic jet
that originates in the innermost regions of the accretion flow.  Of these
two components, the former is thought to dominate if the black hole is
spinning slowly; the frame-dragging of the accretion flow within the
ergosphere can greatly magnify the amplitude of the outflow
\citep{Meier99,Meier01,Nemmenetal07,BensonBabul09}.  The power of the
electromagnetic component rises very steeply at high black hole spins.
When the accretion flow reaches the black hole event horizon, the gas is
expected to drain off the field lines while the field lines, which for all
intents and purposes can be thought of as being anchored on the black
hole's event horizon, establish a structure similar to that first described
by \citet{BlandfordZnajek77}.  In the event that the black hole is
spinning, the winding of the magnetic field lines in the ergosphere will
drive helical twists that manifests as highly collimated, Poynting-flux
dominated jets.  In the picture outlined here, jets arise regardless of
whether the black hole is spinning or not; however, a rapidly spinning
black hole can greatly enhance the total outflow power.

However, it has been argued that the mechanical power of radio jets is
unlikely to represent more than $\sim$10\% of the total energy budget of
all types of AGN \citep{CattaneoBest09}. The remaining 90\% is released
radiatively, generally by AGN with little or no jet activity, but it is as
yet unclear whether this radiation can effectively heat gas on galactic or
cluster scales.

Current structure formation models suggest that feedback from black holes
at redshifts $>$1 is required to establish the shallow gas entropy profiles
in groups \citep[e.g.,][]{McCarthyetal11} and quench star formation in
giant galaxies, so as to produce the distribution of luminosities
  and colours observed today \citep{Boweretal06,Crotonetal06}. It is
widely assumed that radiative ``quasar--mode'' feedback, which should
dominate at this epoch, can effectively heat the ICM via a wind which blows
any cool gas and dust out of the galaxy \citep[e.g.,][]{Hopkinsetal05} and
drives shocks into the ICM.  However this is a largely untested hypothesis,
since powerful quasars are rare at low--redshifts ($z\la1$) where current
instrumentation could resolve such winds, particularly in the dense cluster
environments where their impact could be most easily studied.  The handful
of examples which have been examined using \chandra\ and \xmm\
\citep[e.g.,][]{Belsoleetal07,Siemiginowskaetal10,Russelletal10,Russelletal12}
show no evidence of strong quasar--mode heating, but the difficulty of
identifying disturbed structures or ascertaining the current accretion
state means that the issue is as yet unresolved.

We have chosen to investigate this issue by examining IRAS~09104+4109 (also
known as IRAS~J0913454+405628) a quasar and FR-I radio galaxy located in
the dominant galaxy of the $z$=0.442 galaxy cluster CL~09104+4109 (also
known as MACS J0913.7+4056 and hereafter referred to as CL09). In this
paper we combine new multi-frequency \textit{Giant Metrewave Radio
  Telescope} (\textit{GMRT}) and archival \textit{Very Large Array}
(\textit{VLA}) and \textit{Very Long Baseline Array} (\textit{VLBA})
observations with a reanalysis of the \chandra\ and \xmm\ data, to examine
in detail the interaction between the AGN, radio jets and ICM in CL09.
Section~\ref{sec:literature} describes our current knowledge of the cluster
and AGN, based on previous multiwavelength studies in the literature.  In
Sections~\ref{sec:radioobs} and \ref{sec:obs} we describe the radio and
X--ray observations, and their reduction. The resulting radio images are
shown in Section~\ref{sec:images}, and X-ray imaging and spectroscopic
analysis is described in Section~\ref{sec:Xa}.  Section~\ref{sec:spec}
describes the spectral analysis of the radio source and presents limits on
its age and physical properties.  In Section~\ref{sec:sum} we summarise our
observational results, examine the impact of the radio source on the ICM,
and consider the origin of various structures in the cluster. In
Section~\ref{sec:discuss} we discuss the likely history of the cluster, BCG
and AGN, and the ability of the QSO to balance radiative cooling in the
ICM. Our conclusions are summarised in Section~\ref{sec:conc}.

Throughout the paper we assume a flat cosmology with H$_0$ = 73 km s$^{-1}$
Mpc$^{-1}$, $\Omega_{\lambda}$ = 0.73 and $\Omega_{0}$ = 0.27. For the
redshift of CL09 ($z=0.4418$, Sloan Digital Sky Survey Data Release 3),
this gives a luminosity distance 2372~Mpc, an angular diameter distance
1140~Mpc, and angular scale of 1$^{\prime\prime}$=5.5~kpc. We assume a
galactic hydrogen column of $1.8\times10^{20}$\pcmsq, taken from the
Leiden/Argentine/Bonn Galactic \Hi\ Survey \citep{Kalberlaetal05}.

\subsection{Summary of previous studies}
\label{sec:literature}
IRAS~09104+4109 is a hyper-luminous infrared galaxy \citep[HLIRG,][]{SandersMirabel96}, having an
infrared luminosity $>$10$^{46}$\ergps\
\citep[$>$10$^{13}$\Lsol,][]{Kleinmannetal88}.  The flux density of the
galaxy peaks at $\sim$50~$\mu$m and spectral energy distribution (SED)
modelling suggests that $\sim$70\% of the emission between UV and Sub-mm is
produced by a dust--shrouded active nucleus \citep{Ruizetal10} with a total
bolometric luminosity $\approx(2.3-3.8)\times10^{47}$\ergps\
\citep{Vignalietal11}.  The AGN is thus classed as a type II QSO, with dust
reprocessing almost all its optical and ultraviolet emission into the
infrared \citep{Kleinmannetal88,Vignalietal11,Ruizetal10}.

The remaining $\sim$30\% of the emission from the galaxy arises from
star--formation \citep{Vignalietal11,Ruizetal10}. The galaxy colour profile
becomes significantly bluer within 20~kpc of the centre, and the
star--formation rate (SFR) based on the equivalent width of [O\textsc{ii}]
is 41$\pm$12 \Msolpyr\ \citep{Bildfelletal08}. Modelling of optical and
near-ultraviolet (NUV) photometry suggests that recent star formation with
a probable age of $\approx$70-200~Myr has contributed a small but
significant fraction of the stellar mass in the brightest cluster galaxy
\citep[BCG, probably $>$5\%][]{Pipinoetal09}.

Although the QSO is obscured, the BCG contains relatively little dust,
$<$5$\times$10$^7$\Msol\ \citep{Combesetal11}, with a mean temperature
$\sim$120~K \citep{Kleinmannetal88}.  Based on a detection of CO emission
\citet{Combesetal11} estimate that the BCG contains only
$\sim$3.2$\times$10$^9$\Msol\ of molecular hydrogen, significantly less
than many star--formation dominated ULIRGs \citep[see also][]{Evansetal98}.
It is possible that the star--formation and QSO have consumed or evaporated
much of the cool material in the system, or that the winds from the QSO
have swept the BCG clean.

\begin{figure}
\includegraphics[width=\columnwidth]{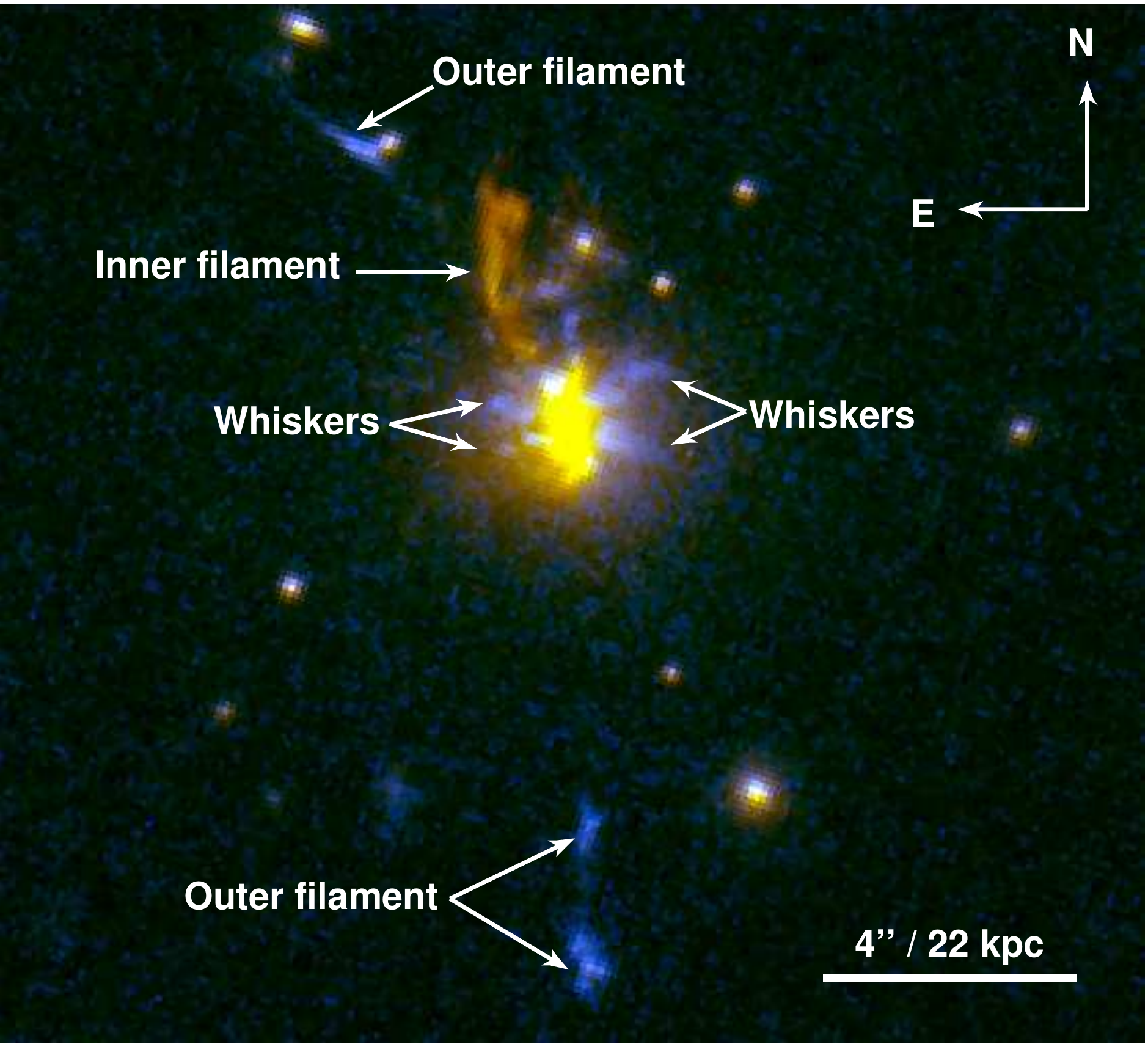}
\caption{\label{fig:HST} False colour \textit{HST} PC2 image of
  IRAS~09104+4109 and its surroundings. An 814W filter image is used for
  the red band, a 622W image for green, and a scaled 622W subtracted from
  the 814W image for the blue band. This combination is chosen to emphasize
  the small--scale filamentary structures referred to in the text.
  Correcting for the redshift of CL09, the 622W and 814W filters have
  central wavelengths (and widths) of $\sim$4300\AA\ ($\sim$635\AA) and
  $\sim$5550\AA\ ($\sim$1070\AA) respectively. The redder colour of the
  inner filament corresponds to strong [O\textsc{iii}] 5007\AA\ emission
  from this structure \protect\citep{Armusetal99}.}
\end{figure}

Ground-based spectrophotometry shows a ``plume'' of [O\textsc{iii}]
emission extending $\sim$5\arcs\ ($\sim$27~kpc) north of the galaxy, with a
velocity only $\sim$100\kmps\ different from that of the nucleus
\citep{CrawfordVanderriest96}. \textit{Hubble Space Telescope}
(\textit{HST}) imaging shows this to be an extended cool gaseous filament
\citep[][labelled ``inner filament'' in
Fig.~\ref{fig:HST}]{Armusetal99} illuminated by spectrally hard
photons from the AGN \citep{CrawfordVanderriest96}, while ground--based
spectroscopy suggests that it contains dust \citep{Tranetal00}.
Other filamentary structures around the BCG include apparent
gaseous filaments north and south of the BCG at radii of 5-9\arcs\
(27-50~kpc, labelled ``outer filaments'') and smaller--scale ``whiskers''
extending 5-7~kpc from the BCG on roughly east-west axes \citep[][see
Fig.~\ref{fig:HST}]{Armusetal99}.  These filaments have
significantly bluer colours than the [O\textsc{iii}] plume, but their
origin is unclear.  Possibilities include material stripped from cluster
galaxies, gas cooling from the hot ICM, or material illuminated by the AGN
through openings in the dust shrouding the BCG.

On arcsecond scales \textit{HST} imaging shows highly polarized emission
north and south of the AGN, with a polarization angle roughly perpendicular
to its extension. These structures are thought to be the optical/UV
scattering cones of the AGN \citep{Hinesetal99}. Ground--based polarimetry
also shows the [O\textsc{iii}] filament to be highly polarized
\citep{Tranetal00}.  Attempts to determine the true opening angle of the
cones and their inclination to the line of sight result in a range of
solutions \citep[half-opening angles 15\degree-40\degree and inclinations
34\degree-50\degree][]{Tranetal00,Hinesetal99}, but the projected
half-opening angle measured from the \textit{HST} images is
35\degree$\pm$10\degree\ \citep{Hinesetal99}.

\citet{HinesWills93}, using VLA 1.4 and 5~GHz observations, found that
IRAS~09104+4109 also hosts a double-lobed radio source
(NVSS~J091345+405630) with straight jets extending $\sim$60~kpc northwest
and southeast of the nucleus, whose radio luminosity places it in the FR~I-
FR~II transition region. Its spectral
index\footnote{The radio spectral index $\alpha$ is defined according to
  $S_{\nu}\propto \nu^{-\alpha}$, where $S_{\nu}$ is the flux density at
  the frequency $\nu$.} is $\alpha_{1.4-5~GHz}\approx1.4$, fairly steep for
a radio galaxy .  The jet axis lies just outside the ionisation cones of
the AGN, and it has been suggested that this misalignment, along with the
brightness of the radio core (unresolved in 6cm and 20cm VLA observations)
may indicate that AGN has recently changed orientation, with a new jet
aligned close to the line of sight producing the bright core via
relativistic beaming \citep{HinesWills93,Hinesetal99,Armusetal99}.

\rosat\ HRI observations of the CL09 cluster showed extended X-ray emission
centred on the BCG, with a central hole $\sim$4\arcs\ ($\sim$20~kpc) across
\citep{FabianCrawford95}. \rosat\ imaging and \asca\ spectral analysis both
indicate the presence of a strong cooling flow in the cluster core
\citep{FabianCrawford95,CrawfordVanderriest96}. More recent \chandra\
observations have confirmed the presence of a strong cool core with a short
cooling time (2.1~Gyr within $\sim$50~kpc) and shown that the hole is one
of two cavities located northwest and southeast of the core along the jet
axis \citep{HlavacekLarrondoetal11,Iwasawaetal01}.  The enthalpy of these
cavities is more than sufficient to balance radiative losses within the
cooling region \citep[][hereafter HL11]{HlavacekLarrondoetal11}.

\section{Radio Observations and Data Reduction}
\label{sec:radioobs}

\subsection{GMRT Observations}
\label{sec:gmrt}

CL09 was observed using the \gmrt\, at 240 MHz, 330 MHz, 610 MHz (project
15SRC01) and 1.28 GHz (project 17$_{-}$050) in a full-synthesis run of
approximately 11 hours at 1.28 GHz and 9 hours at each of the other
frequencies (including calibration overheads). Details on these
observations are summarised in Table \ref{tab:gmrtobs}, which reports:
observing date, frequency and total bandwidth, total time on source,
full-width half maximum (FWHM) and position angle (PA) of the full array,
and rms level (1$\sigma$) at full resolution.

The 330 MHz and 1.28 GHz data were recorded using the upper and lower side
bands (USB and LSB, respectively) simultaneously, which provide a total
observing bandwidth of 32 MHz. The 240 MHz and 610 MHz observations were
carried out in simultaneous mode with 32 MHz bandwidth (USB+LSB) at 610 MHz
and 8 MHz at 240 MHz. The data were collected using the default
spectral-line mode with 128 channels for each band, resulting in a spectral
resolution of 125 kHz per channel. The data sets were calibrated and
reduced using the NRAO\footnote{National Radio Astronomy Observatory.}
Astronomical Image Processing System (AIPS) package. The data were
initially inspected using the task SPFLG to identify and remove bad
channels and visibilities affected by radio frequency interference (RFI).
The data were then calibrated. The flux density scale was set using
amplitude calibrators (3C48, 3C147 and 3C286), observed at the beginning
and at the end of the observing run, and the \citet{PerleyTaylor99}
extension of the \citep{Baarsetal77} scale. The source 0834+555 was used as
phase calibrator at all frequencies.

The bandpass calibration was carried out using the flux density
calibrators. A central channel free of RFI was used to normalise the 
bandpass for each antenna. After bandpass calibration, the central 
84 (48) channels were averaged to 6 channels of $\sim$2 MHz (1 MHz) at
330 MHz, 610 MHz and 1.28 GHz (240 MHz) to reduce the size of the 
data set and, at the same time, to minimise the bandwidth smearing 
effects within the primary beam of the \gmrt\, antenna. 

After further careful editing of the in the averaged data, a number of
phase-only self-calibration cycles and imaging were carried out for each
data set. The large field of view of the \gmrt\, required the
implementation of wide-field imaging technique in each step of the data
reduction, to account for the non-planar nature of the sky. The USB and LSB
were calibrated separately. The final data sets were further averaged from
6 channels to 1 single channel\footnote{Bandwidth smearing is relevant only
  at the outskirts of the wide field, and does not significantly affect the
  region presented and analysed in this paper.}, and then combined together
to produce the final images. These were corrected for primary beam pattern
of the \gmrt\, antenna using PBCOR in AIPS. The rms noise level (1$\sigma$)
achieved in the final full resolution images are summarised in Table
\ref{tab:gmrtobs}.

The average residual amplitude errors on each individual antenna are
$\ltsim$ 5\% at 610 MHz and 1.28 GHz and $\ltsim$ 8\% at 240 MHz and 330
MHz \citep[e.g.,][]{Chandraetal04}. Therefore, we can conservatively assume
that the absolute flux density calibration is within 5$\%$ at 610 MHz and
1.28 GHz and 8\% at lower frequency.

\begin{table*}
\caption[]{Details of the {\em GMRT} radio observations}
\begin{center}
\begin{tabular}{ccccccc}
\hline\noalign{\smallskip}
 Observation & Frequency & Bandwidth & Integration  & FWHM, PA  & rms      \\
 date & (MHz)          &  (MHz)        &  time (min)      &
	    (full array, $^{\prime \prime} \times^{\prime \prime}$,
            $^{\circ}$) & (mJy beam$^{-1}$) \\
\noalign{\smallskip}
\hline\noalign{\smallskip}
Dec 2008 & \phantom{0}240 $^{a}$& 8 & 390 & $19.9\times19.3$, $-7$ & 1.3 \\
Mar 2009 & 327 & 32& 280&  $11.1\times9.1$, 17 & 0.20 \\
Dec 2008 & \phantom{0}610 $^{a}$& 32 & 390 & $6.1\times5.2$, $-73$  & 0.08 \\
Dec 2009 & 1280 & 32 & 390 & 2.8$\times$2.1, 53 & 0.02 \\
\hline{\smallskip}
\end{tabular}
\end{center}
\label{tab:gmrtobs}
Notes to Table \ref{tab:gmrtobs} -- $a:$ Observed in dual 240/610 MHz mode.
\end{table*}

\begin{table*}
\caption[]{Summary of the VLA archive observations}
\begin{center}

\begin{tabular}{cccccccc}
\hline\noalign{\smallskip}
Project & Array & Observation  & Frequency & Bandwidth & Integration  & FWHM,
PA  &   rms    \\ 
            &  & date  & (GHz)            &  (IF1/IF2, MHz)        &  time (min)      &
	    (full array, $^{\prime \prime} \times^{\prime \prime}$, $^{\circ}$) &
            ($\mu$Jy b$^{-1}$) \\
\noalign{\smallskip}
\hline\noalign{\smallskip}
AH406 & AnB &  Jul 1990       &  8.41/8.46 &  50 & 40  & 0.9$\times$0.7, 72&
20 \\
AT211 & A&  Apr 1998      &   1.47/1.67 & 25 & 110 & 1.3$\times$1.1, $-64$ & 25 \\
AT211 & B & Aug 1998       &  1.47/1.67 & 25 &  40  &  4.6$\times$4.1, $-40$ & 40\\
AT211 & A & Apr 1998       &   4.64/4.89 & 50 & 30 & 0.4$\times$0.4, $-71$& 50 \\
AT211  &B&  Aug 1998      &   4.64/4.89 & 50  & 90 & 1.4$\times$1.4, $-33$& 25 \\
AT211  & C&  Apr 1998      &   4.64/4.89 & 50  & 45 & 7.2$\times$4.1, $81$ & 40 \\
\hline{\smallskip}
\end{tabular}
\end{center}
\label{tab:vla}
\end{table*}

\subsection{VLA Archive Data}
\label{sec:vla}

We extracted all observations of CL09 useful for the analysis presented in
this paper from the \vla\, public archive and re-analysed them.  The
observing details are provided in Table \ref{tab:vla}, which shows the
project code and array configuration in the first two columns; the other
columns provide the same information as Table \ref{tab:gmrtobs}.

Calibration and imaging were carried out using AIPS, following the standard
procedure, i.e., Fourier-Transform, Clean and Restore.  We applied
phase-only self-calibration to remove residual phase variations and improve
the quality of the images. Correction for the primary beam attenuation was
applied to the final images using the task PBCOR in AIPS. All flux
densities are on the \citet{PerleyTaylor99} extension of the
\citet{Baarsetal77} scale. Average residual amplitude errors are
$\ltsim5$\% at all frequencies.

\subsection{VLBA Archive Data}
\label{sec:vlba}

CL09 was observed at 1.4 GHz by the {\em VLBA} on 2003 March 13 (project
BH0110) for approximately 12 hours in phase referencing, with a recording
bandwidth of 8 MHz. We extracted the calibrated data from the {\em VLBA}
archive, processed with the {\em VLBA} data calibration pipeline
\citep{Sjouwermanetal05} implemented in AIPS. The data set was further
self-calibrated in phase with the AIPS task CALIB to reduce phase
fluctuations, and used to produce the final image with a resolution of 8.3
mas $\times$ 6.3 mas.  The rms noise level measured on the image plane is
25 $\mu$Jy beam$^{-1}$.

\section{X-ray Observations and Data Reduction}
\label{sec:obs}

CL09 was observed during \chandra\ cycle 10, on 2009 January 06 for
$\sim$77~ks (ObsID 10445), with the ACIS-I instrument operating in very
faint telemetry mode. A summary of the \chandra\ mission and
instrumentation can be found in \citet{Weisskopfetal02}. The data were
reduced and analysed using \textsc{ciao} 4.3 and CALDB 4.4.3 following
techniques similar to those described in \citet{OSullivanetal07} and the
\chandra\ analysis
threads\footnote{http://asc.harvard.edu/ciao/threads/index.html}. The
observation did not suffer from significant background flaring, and the
final cleaned exposure time was 73.3~ks.

Point sources were identified using the \textsc{wavdetect} task, with a
detection threshold of 2.38$\times$10$^{-7}$, chosen to ensure that the
task detects $\leq$1 false source in the ACIS-I field, working from a
0.3-7.0 keV image and exposure map. All point sources were excluded, except
in some cases the source corresponding to the cluster--central AGN.
Spectra were extracted using the \textsc{specextract} task. Spectral
fitting was performed in \textsc{XSpec} 12.7.0e. Abundances were measured
relative to the abundance ratios of \citet{GrevesseSauval98}. A galactic
hydrogen column of 0.018$\times10^{22}$\pcmsq\ and a redshift of 0.44 were
assumed in all fits. Spectra were grouped to 20 counts per bin, and counts
at energies outside the range 0.5-7.0 keV were generally ignored during
fitting.

Background spectra were drawn from the standard set of CTI-corrected ACIS
blank sky background events files in the \chandra\ CALDB. The exposure time
of each background events file was altered to produce the same 9.5-12.0 keV
count rate as that in the target observation. Very faint mode background
screening was applied to both source and background data sets.

\xmm\ has also observed CL09, for $\sim$14~ks (ObsId 0147671001). This
observation is described in detail in \citet{Piconcellietal07} and a
detailed summary of the \xmm\ mission and instrumentation can be found in
\citet[and references therein]{Jansenetal01}. To provide a comparison with
\chandra\ image analysis, we reduced this data using \textsc{sas} v11.0.1
following the methods described in \citet{OSullivanetal11c}.  The EPIC
instruments were operated in full frame mode, with the medium optical
blocking filter. Periods including background flaring, when the total count
rate deviated from the mean by more than 3$\sigma$, were excluded, leaving
useful exposures of 12.0~ks (MOS1), 12.3~ks (MOS2) and 8.4~ks (pn). Point
sources were identified using \textsc{edetect$\_$chain}, and regions
corresponding to the 85 per cent encircled energy radius of each source
(except that at the peak of the diffuse emission) were excluded.

\section{Radio analysis}
\label{sec:images}

\begin{figure}
\centering
\includegraphics[width=\columnwidth]{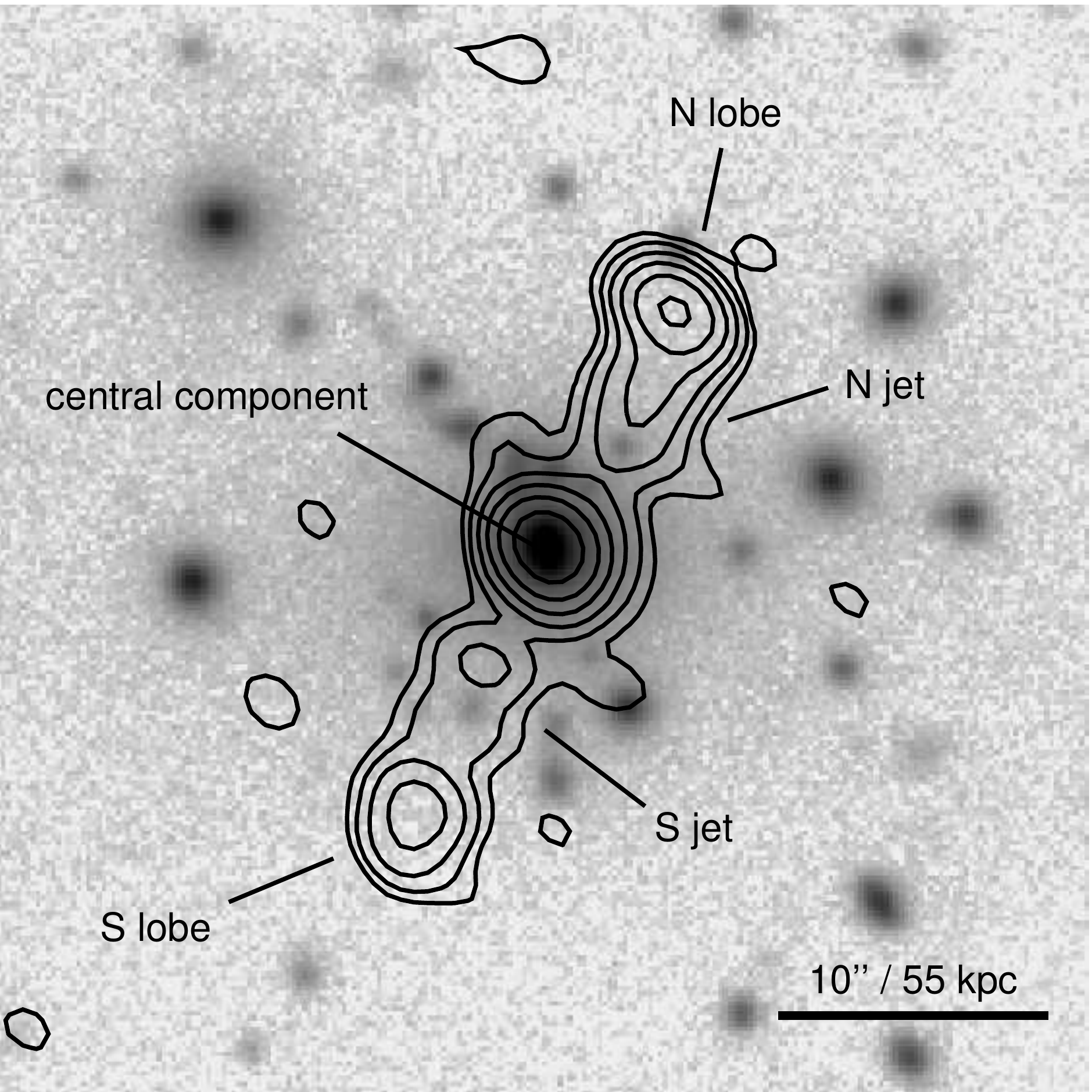} 
\caption{{\em GMRT} contours at 1.28 GHz, overlaid on
  a CFHT r'--band image. The restoring 
beam is $2.8^{\prime \prime}\times2.1^{\prime\prime}$, p.a. 53$^{\circ}$. The rms
noise level is 1$\sigma$=20$\mu$Jy. Contour levels are spaced by a
factor of 2 starting from +3$\sigma$. Labels indicate the individual
components of the radio source.}
\label{fig:1.28}
\end{figure}

\begin{figure*}
\centering
\includegraphics[width=5.8cm]{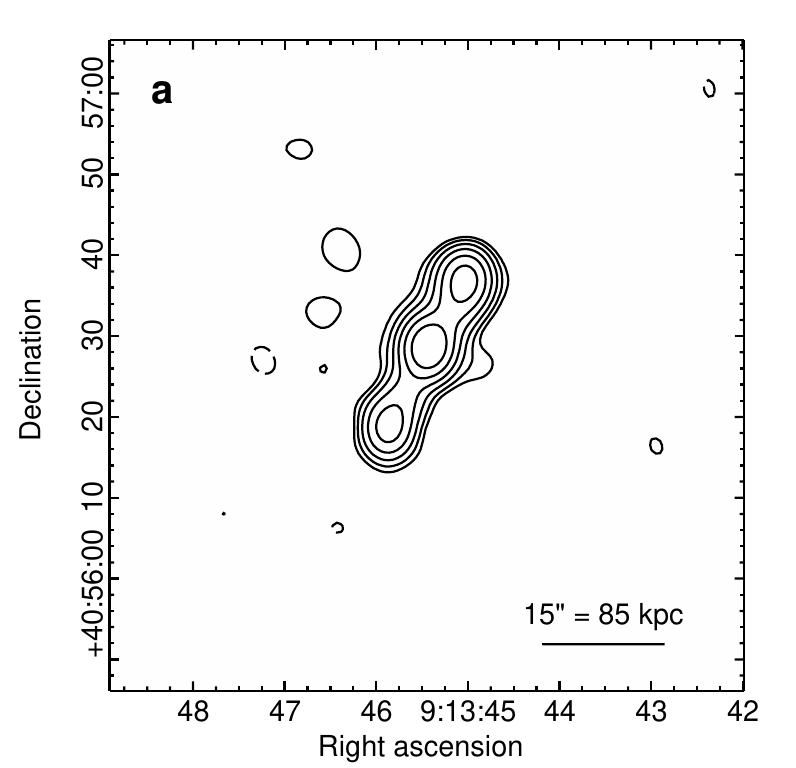}
\includegraphics[width=5.8cm]{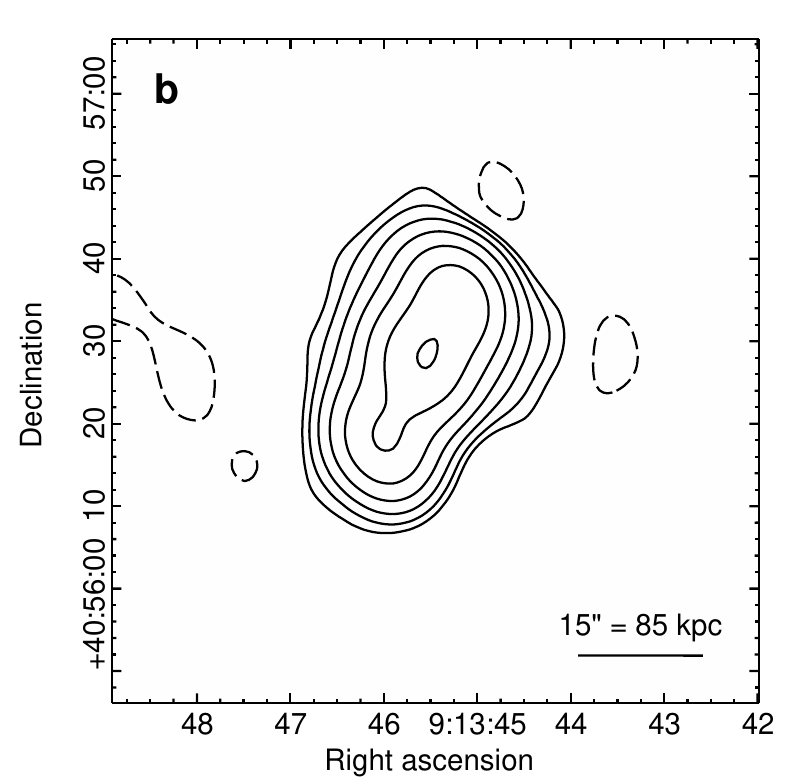}
\includegraphics[width=5.8cm]{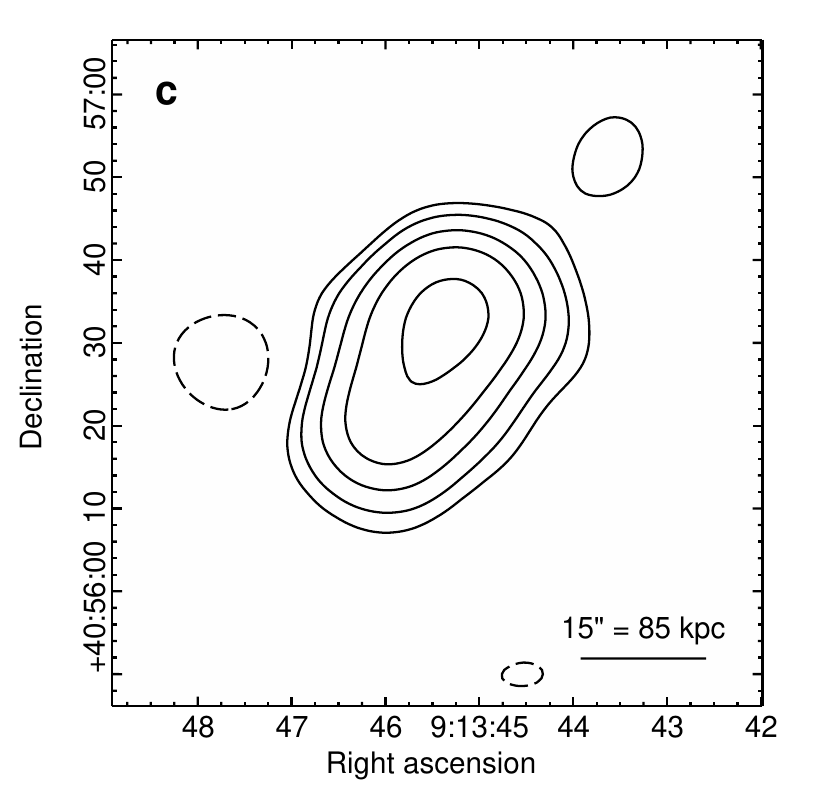}
\caption{{\em GMRT} images at 610 MHz ({\em a}), 330 MHz ({\em b}) and
  240 MHz ({\em c}). The resolution and rms noise 
are {\em a)}  $5.0^{\prime\prime}\times4.0^{\prime\prime}$,
p.a. 0$^{\circ}$, 1$\sigma$=80 $\mu$Jy beam$^{-1}$; 
{\em b)} $10.5^{\prime\prime}\times8.1^{\prime\prime}$,
p.a. 25$^{\circ}$; 1$\sigma$=0.2 mJy beam$^{-1}$; {\em c)} 
$13.5^{\prime\prime}\times11.5^{\prime\prime}$,
p.a. 64$^{\circ}$; 1$\sigma$=1.3 mJy beam$^{-1}$. 
Contours are spaced by a factor of 2 starting from at +3$\sigma$. 
Dashed contours correspond to the $-3\sigma$ level.}
\label{fig:low_gmrt}
\end{figure*}

The 1.28 GHz \gmrt\, image at the resolution of $2.8^{\prime
  \prime}\times2.1^{\prime\prime}$ is presented in Fig.~\ref{fig:1.28}, as
contours overlaid on the r'--band optical image from the Canada--France--Hawaii Telescope (CFHT). The image shows a double-lobe radio source with two fairly
straight and weak jets and a dominant, compact component at the location of
the cD galaxy. Fig.~\ref{fig:low_gmrt} presents the \gmrt\, images at
610 MHz, 330 MHz and 240 MHz. While the source is resolved into a triple at
the 5$^{\prime\prime}$-resolution of the 610 MHz image, it is not spatially
resolved in the images at 330 MHz and 240 MHz (FWHM$\sim$11$^{\prime
  \prime}$ and $\sim$14$^{\prime \prime}$, respectively).  The source has
an angular size of $\sim 27^{\prime \prime}$, corresponding to a linear
size of $\sim$150 kpc, fairly consistent at all frequencies and
resolutions.

\begin{figure*}
\centering
\includegraphics[width=16cm]{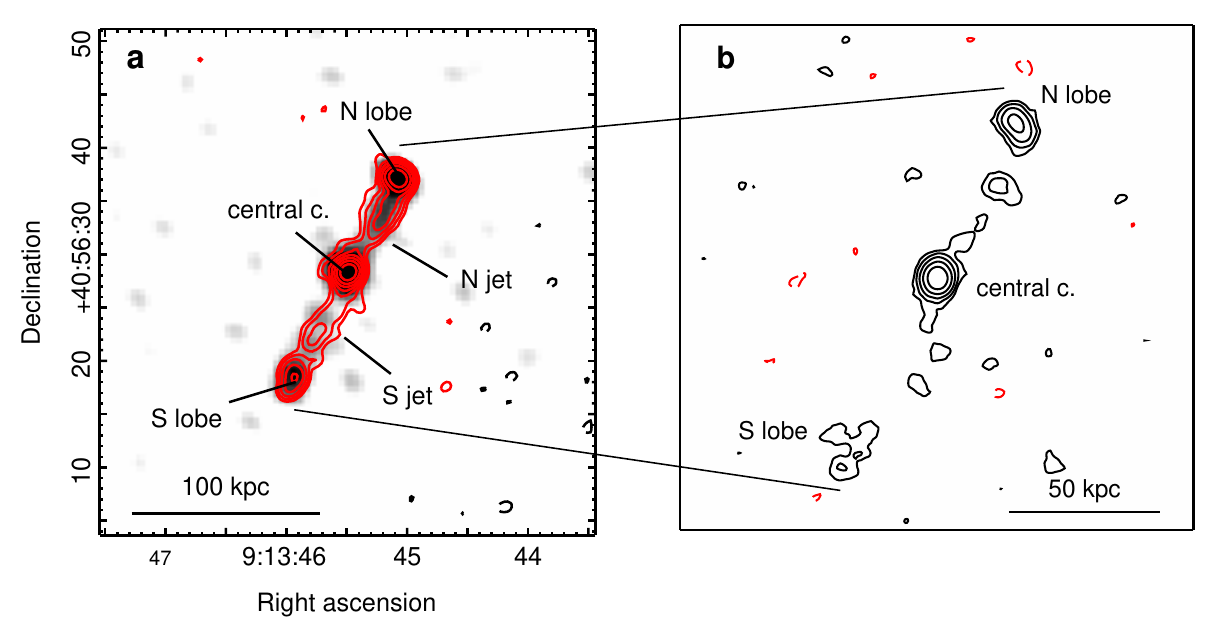}
\hspace{-0.5cm}
\caption{{\em a)}: 
{\em VLA} 1.5 GHz contours, obtained from the combined A+B array data set,
overlaid on the {\em GMRT} 1.28 GHz image (same as
Fig.~\ref{fig:1.28}). The resolution of the 1.5 GHz image is 
$1.5^{\prime\prime}\times1.3^{\prime\prime}$, p.a. $-$68$^{\circ}$). 
Red contours start at +3$\sigma$=45 $\mu$Jy
beam$^{-1}$ and then scale by a factor of 2. Black dashed contours 
correspond to the $-3\sigma$ level.
{\em b)}: {\em VLA} 4.8 GHz contours from the combined A+B 
array data set. The resolution is
$1.3^{\prime\prime}\times1.2^{\prime\prime}$, p.a. 51$^{\circ}$ and
1$\sigma$=20 $\mu$Jy beam$^{-1}$.  Black contours start at +3$\sigma$
and then scale by a factor of 2. Red dashed contours correspond to
the $-3\sigma$ level.}
\label{fig:vla}
\end{figure*}

Fig.~\ref{fig:vla}a shows the \vla\, image at 1.5 GHz, obtained from the
combination of the A- and B-array data sets \citep[Table \ref{tab:vla}; see
also][]{HinesWills93}. The source morphology in this image is very similar
to Fig.~\ref{fig:1.28}.  The combined A+B-array contour image at 4.8 GHz is
presented in Fig.~\ref{fig:vla}b. Here, the brightest features are the
central component and northern lobe; the S lobe is very faint and jets are
barely visible.  The central component and N lobe are also detected in the
$\sim 1^{\prime \prime}$-resolution image at 8.4 GHz (not shown here; Table
\ref{tab:vla}).

\begin{figure}
\centering
\includegraphics[width=\columnwidth]{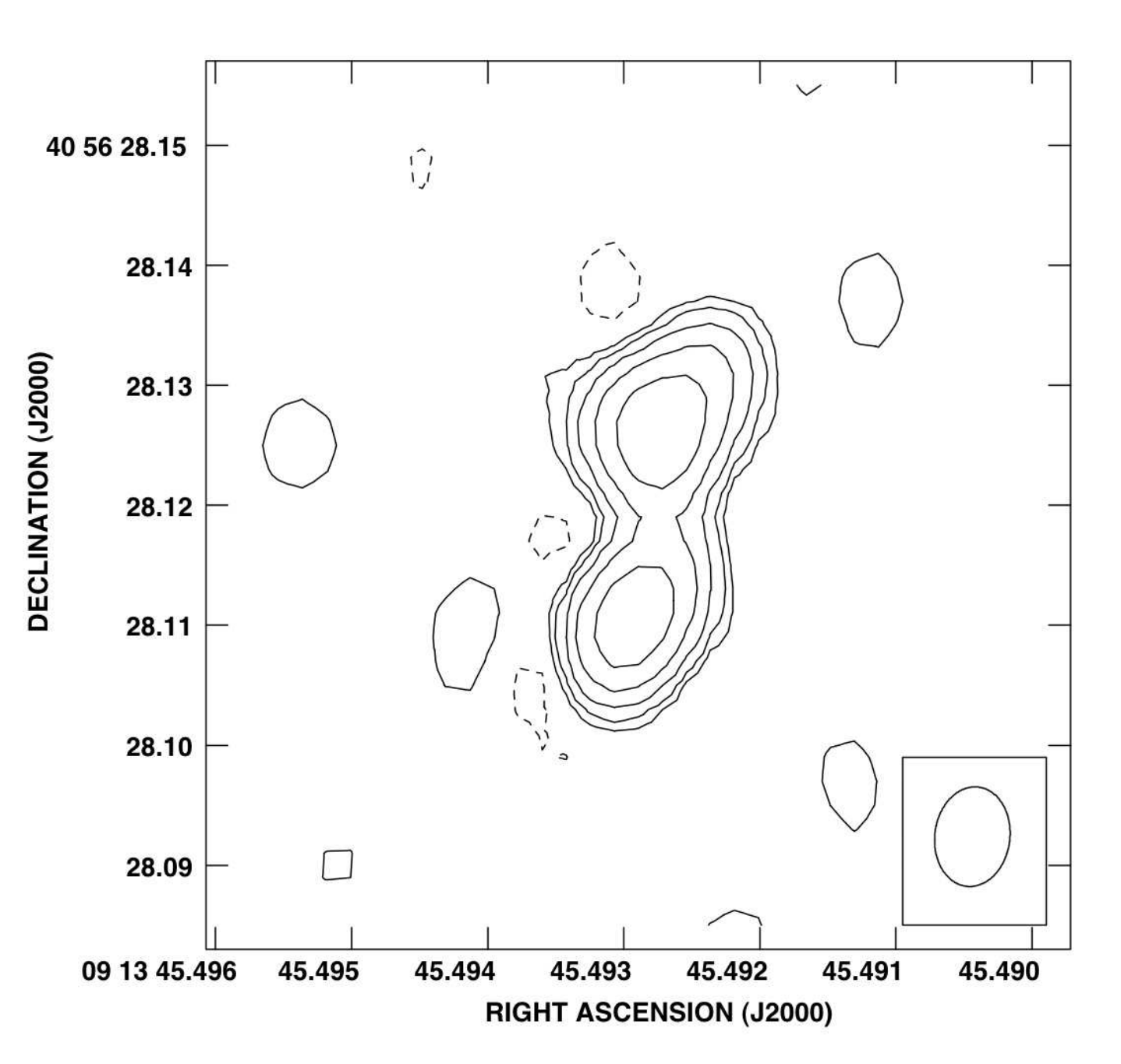} 
\caption{{\em VLBA} image at 1.4 GHz of CL09. The restoring 
beam is 
$0.0083^{\prime \prime}\times0.0063^{\prime\prime}$, p.a. $-$7$^{\circ}$. The rms
noise level is 1$\sigma$=25$\mu$Jy. The peak flux density is 2.3 mJy beam$^{-1}$.
Contour levels are 75$\times$[-1 (dashed), 1, 2, 4, 8, 16] $\mu$Jy
beam$^{-1}$. }
\label{fig:vlba}
\end{figure}

Fig.~\ref{fig:vlba} zooms on the central region of CL09, by presenting the
{\em VLBA} image at 1.4 GHz. At VLBA resolution ($8\times6$ mas), the
central component is resolved into a quite symmetric double of $\sim$35 mas
size (i.e., $\sim$0.2 kpc). No clear core component is visible. The double
accounts for a total flux density of 5.7 mJy, which is in agreement with
5.9$\pm$0.3 mJy measured for the central component in the \vla\, image at
1$^{\prime \prime}$-resolution (Table \ref{tab:vla}). This implies that no
further, extended structure is present in the central region in addition to
the {\em VLBA} double.

The radio flux densities of CL09, obtained from the \gmrt\, and \vla\,
images presented above, are summarised in Table \ref{tab:total}, along with
the associated uncertainties. The flux density at 4.8 GHz was measured on
the C-array image (not shown here; see Table~\ref{tab:vla}). At 8.4~GHz we
measured a lower limit on the flux, since only the central component and N
lobe were detected. Inspection of the VLSS image at 74 MHz suggests the
presence of a weak, unresolved source at the position of CL09.  The source
is not listed in the VLSS catalogue, implying that its flux density at 74
MHz is less then the $5\sigma$ local rms level of the survey
(1$\sigma$=100~mJy/beam, with FWHM= 80\arcs$\times$80\arcs). We therefore
assume an upper limit of 500 mJy for the flux density of CL09 at 74 MHz,
also reported in Table~\ref{tab:total}. We also include two values from the
literature; the 151~MHz measurement from the 7C survey \citep{Rileyetal99},
corrected to our adopted flux scale \citep{Helmboldtetal08}, and a 15~GHz
measurement made using the Arcminute Microkelvin Imager
\citep{HlavacekLarrondoetal11}. The 1.5~GHz radio power of the source,
1.1$\times$10$^{25}$~W~Hz$^{-1}$ is consistent with the transition between
FR I and FR II sources \citep{BridlePerley84,OwenLaing89,OwenWhite91} as is
often the case with optically luminous cluster and group--central radio
galaxies.

\begin{table}
\caption[]{Radio properties of NVSS~J091345+405630}
\begin{center}
\begin{tabular}{lc}
\hline
Frequency, $\nu$ & $S_\nu$ (mJy) \\
\hline
74~MHz   & $<$500$^{a}$ \\
151~MHz  & 251$\pm$9$^{b}$ \\
240~MHz  & 154.0$\pm$14 \\
327~MHz  & 99.7$\pm$8.0 \\
610~MHz  & 42.6$\pm$2.1 \\
1.28~GHz & 17.6$\pm$0.9 \\
1.5~GHz  & 14.5$\pm$0.7 \\
4.8~GHz  & 3.5$\pm$0.2$^{c}$ \\
8.4~GHz  & $>$1$^{d}$ \\
15~GHz   & 0.80$\pm$0.04$^{e}$ \\
\hline
$\alpha_{\rm 151~MHz-15~GHz}$ & 1.25$\pm$0.01 \\
\hline
\end{tabular}
\end{center}
Notes to Table \ref{tab:total} -- 
$a:$ 5$\sigma$ upper limit from the VLSS;
$b:$ from the 7C survey \citep{Rileyetal99}, converted to our adopted flux density scale using the
conversion factor listed by \citet{Helmboldtetal08};
$c:$ measured on the C-array image;
$d:$ lower limit since only the core and N lobe are detected in the 8.4~GHz
image;
$e:$ Arcminute Microkelvin Imager measurement reported by \citet{HlavacekLarrondoetal11}.
\label{tab:total}
\end{table}

\begin{table*}
\caption[]{Radio properties of the individual components of CL09}
\begin{center}
\begin{tabular}{lccccccc}
\hline\noalign{\smallskip}
Component 
&   $S_{1.28 \, \rm GHz}$ & $S_{1.5
  \, \rm GHz}$ & $S_{4.8 \, \rm GHz}$ &$S_{8.4 \, \rm GHz}$  & 
$\alpha_{\rm obs}$ & $\alpha_{\rm fit}$  \\
                  & (mJy) & (mJy)  &(mJy) & (mJy) & &  \\
\noalign{\smallskip}
\hline\noalign{\smallskip}
Central component  & 7.06$\pm$0.35 & 5.93$\pm$0.30 &
1.51$\pm$0.08 & 0.63$\pm$0.04 &  1.28$\pm$0.04 & 1.23$\pm$0.10 \\
N lobe & 4.19$\pm$0.21 & 3.57$\pm$0.18 & 0.75$\pm$0.04
& 0.38$\pm$0.03 &  1.27$\pm$0.05 & 1.28$^{+0.11}_{-0.09}$ \\
S lobe  & 1.55$\pm$0.08& 1.35$\pm$0.07 &
0.46$\pm$0.03 & $-$ & 0.94$\pm$0.06 & 0.90$^{+0.20}_{-0.11}$  \\
N jet  & 3.45$\pm$0.22 & 2.48$\pm$0.18 & 0.46$\pm$0.03 &$-$ & 1.56$\pm$0.06 & 1.44$^{+0.06}_{-0.09}$\\
S jet  & 1.35$\pm$0.07 & 1.17$\pm$0.06 & 0.32$\pm$0.02 & $-$ & 1.12$\pm$0.06 & 1.07$^{+0.20}_{-0.11}$\\
\hline{\smallskip}
\end{tabular}
\end{center}
\label{tab:comp}
\end{table*}

\section{X-ray analysis}
\label{sec:Xa}

\subsection{X-ray images}
\label{sec:Xims}
\begin{figure}
\includegraphics[width=\columnwidth]{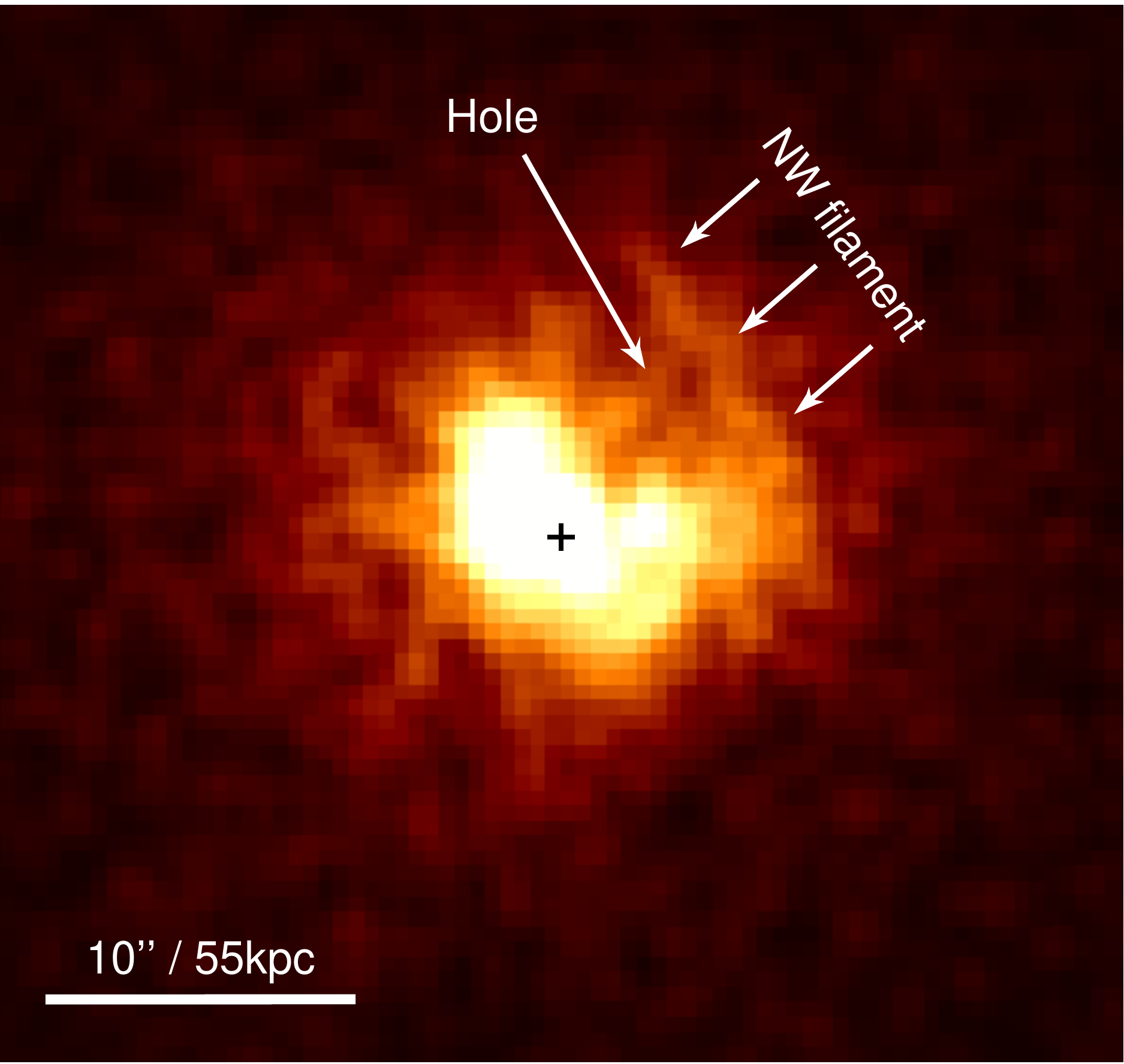}
\caption{\label{fig:XSB}Unbinned 0.3-3~keV exposure corrected \chandra\ image of the
  core of CL09, smoothed with a Gaussian of width 2 pixels
  ($\sim$1\arcs). The black cross marks the optical centre of the BCG.}
\end{figure}

We initially examined the structure of the ICM using the exposure corrected
0.3-3~keV image of CL09 shown in Fig.~\ref{fig:XSB}. As noted by
\citet{Iwasawaetal01}, the ICM emission is brightest to the northeast, west
and southwest of the core, with a depression to the northwest. This hole
was first observed in \rosat\ HRI data
\citep{FabianCrawford95,CrawfordVanderriest96}. The hole is bounded on its
northwest side by a linear filamentary structure. To determine the
significance of the filament, we placed $\sim$1.5\arcs$\times$7\arcs\ box
regions on the filament and immediately to either side, and determined the
number of counts in each region. The exposure corrected 0.3-3~keV surface
brightness of the filament is 2$\sigma$ greater than the region inside (in
the hole) and 3.6$\sigma$ greater than the surface brightness immediately
outside.

As well as the northwest hole, HL11 identify a second cavity to the
southeast.  To provide a clearer view of the structures in the ICM, we
modelled the surface brightness distribution of the cluster in the
\textsc{ciao sherpa} package using an elliptical $\beta$-model, with an
additional constant component to represent the background. A monoenergetic
exposure map with energy 1.39~keV (chosen to match the mean photon energy
of the 0.3-3~keV image) was used to correct for exposure variations across
the source. The effect of the PSF is probably minimal and was therefore
neglected. The best fitting model had core radius 4.42\arcs$\pm$0.11
(24.3$\pm$0.6~kpc), $\beta$=0.578$\pm$0.004, ellipticity 0.136$\pm$0.010,
and position angle 9.3\degree$\pm$2.2, indicating that the major axis of
the ICM runs almost east-west. All parameters of the $\beta$-model were
allowed to vary freely in the fit, including the central position.
Inclusion of a second $\beta$-model had no significant effect on the model
parameters, the second component primarily modelling emission in the core
of the BCG.

\begin{figure}
\includegraphics[width=\columnwidth]{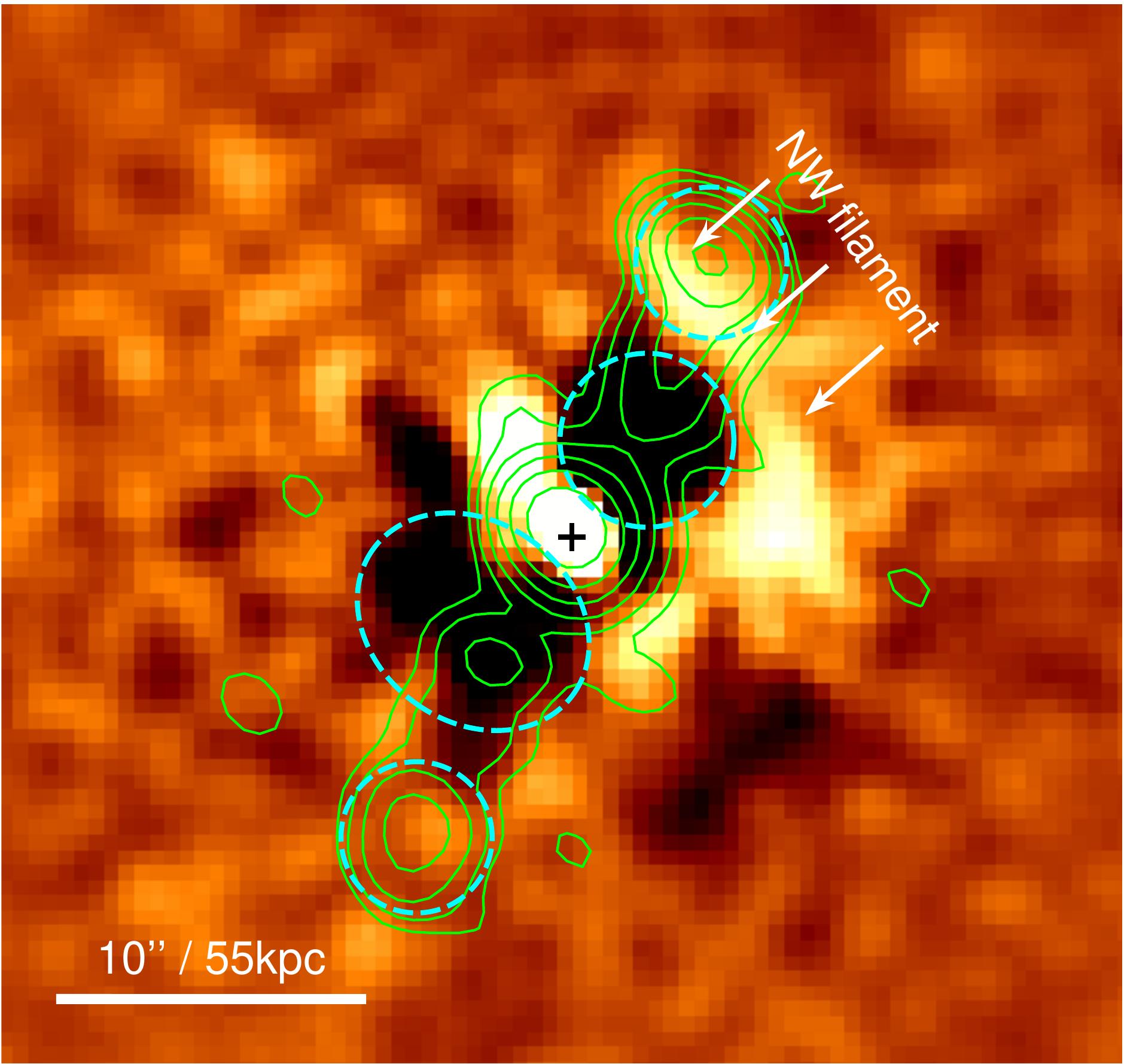}
\caption{\label{fig:resid1}\chandra\ 0.3-3~keV residual image after
  subtraction of the best fitting elliptical $\beta$-model, smoothed with a
  Gaussian of width 2 pixels ($\sim$1\arcs). The northwest filament and
  optical galaxy centre are labelled as in Fig.~\ref{fig:XSB}. Dashed cyan
  circles or ellipses indicate regions used to estimate cavity properties.
  1.28~GHz GMRT contours are overlaid (same as Fig.~\ref{fig:1.28}).}
\end{figure}

\begin{figure}
\includegraphics[width=\columnwidth]{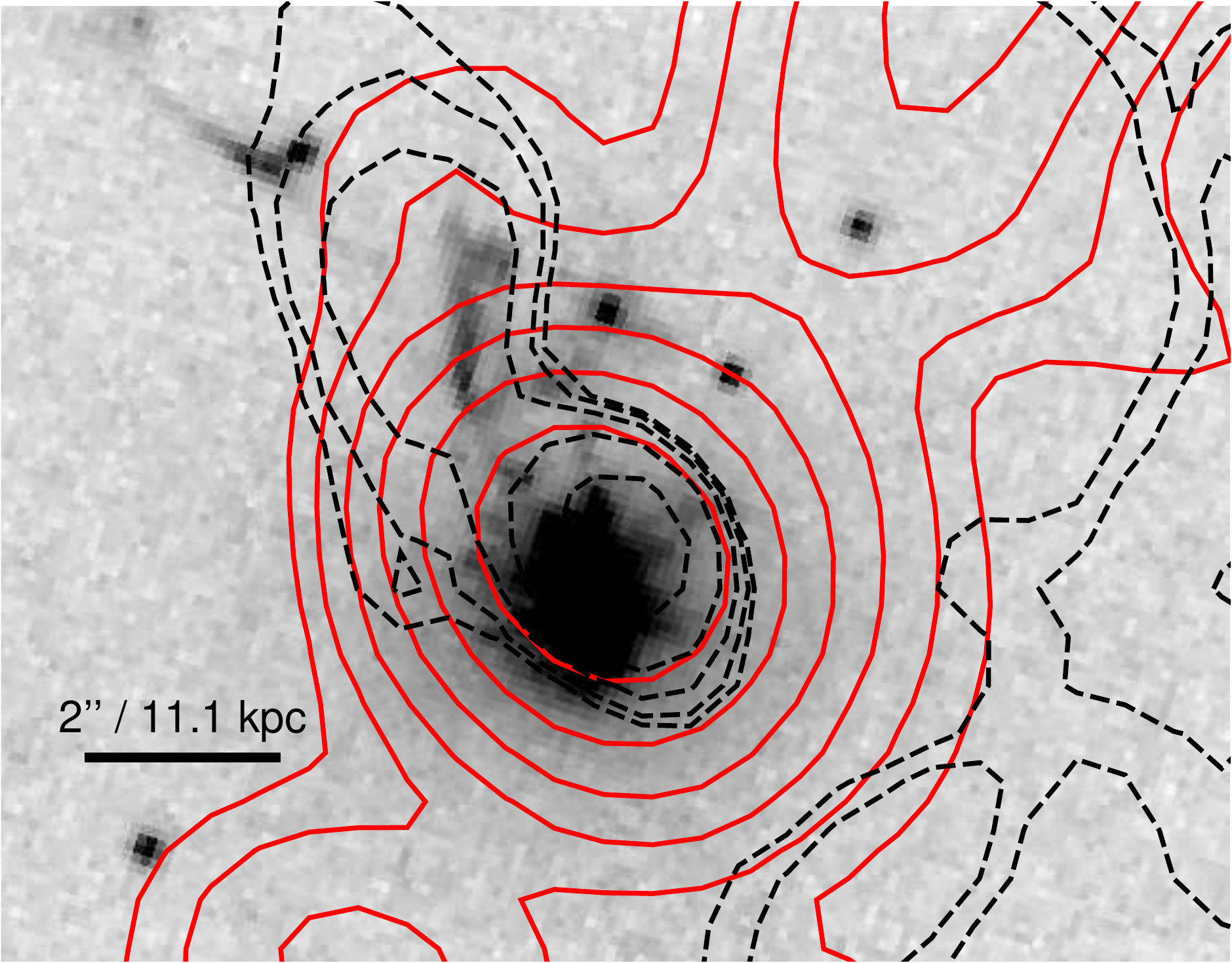}
\caption{\label{fig:spur} \textit{HST} 622W image of the core of CL09,
  overlaid with 1.28~GHz radio contours (solid red, as in
  Fig.~\ref{fig:1.28}) and contours of smoothed positive X--ray residuals
  (dashed black, starting at 0.5 counts/pixel and increasing in steps of
  factor 2).}
\end{figure}

Fig.~\ref{fig:resid1} shows the residual image created by subtracting the
best-fitting model from the 0.3-3~keV image, smoothed with a 2\arcs-width
Gaussian to bring out small-scale features. A central bright source is
visible coincident with the BCG core and inner component of the radio
source. There is a bright extension out to $\sim$5.5\arcs\ northeast,
almost perpendicular to the radio jet axis, which is correlated with the
brightest of the gaseous filaments north of the BCG and with a small spur
of 1.28~GHz radio emission (see Fig.~\ref{fig:spur}). However the
strongest features are two dark regions of negative residuals to northwest
and southeast, roughly coincident with the jet axis, but not extending as
far as the radio lobes. These are the cavities, where thermal ICM plasma
has presumably been driven out by the relativistic plasma of the radio
jets, leaving a deficit in X-ray emission. The northwestern cavity
corresponds to the hole seen in the raw image, and has bright residuals
extending around its western and northern sides, including the northwest
filament. These bright features may arise from ICM gas pushed out of the
cavity.  The southeast cavity is less well-correlated with the jet
structure, extending to the east beyond the radio emission. There is also a
weak negative residual to the southwest of the central component, which is
uncorrelated with the jet.  Our estimates of cavity size and position
differ somewhat from those of HL11. We discuss these differences (and the
physical properties of the cavities) in Section~\ref{sec:cav}.

\begin{figure*}
\includegraphics[width=\textwidth]{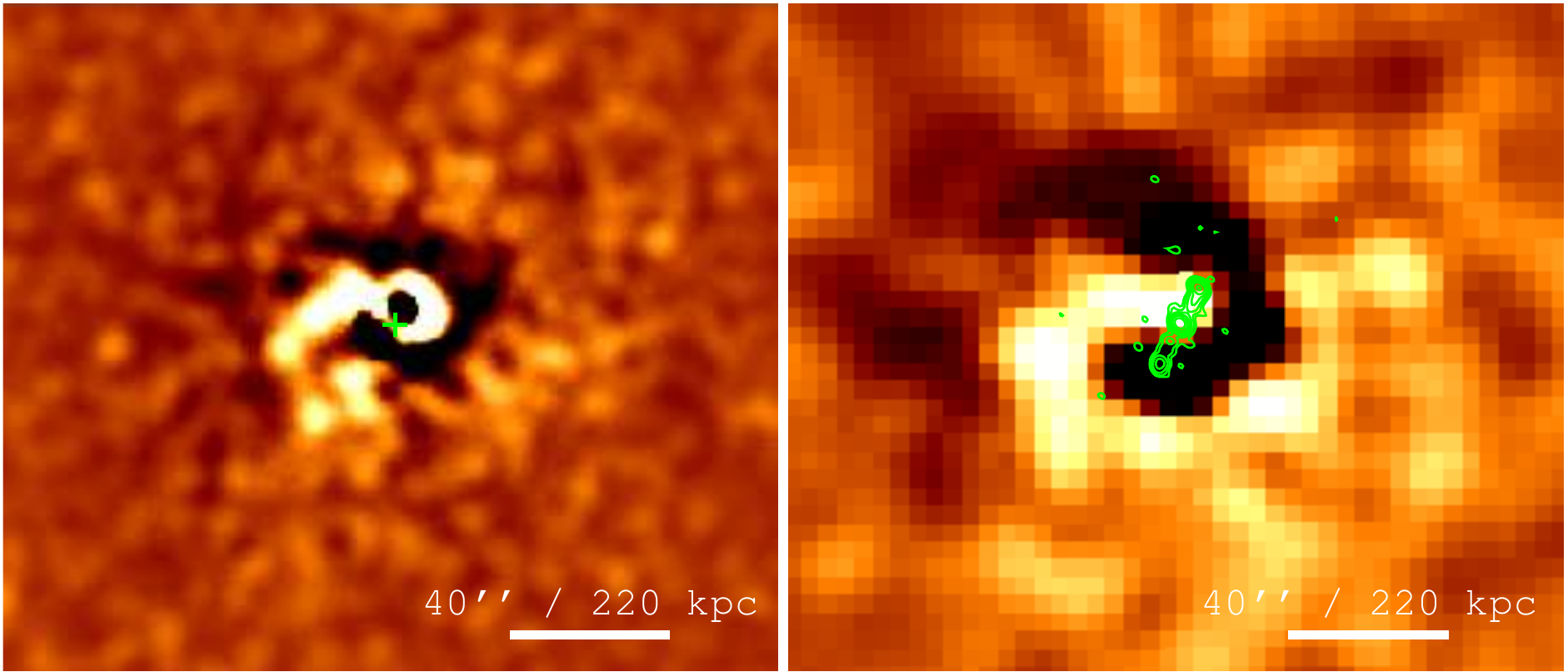}
\includegraphics[width=\textwidth]{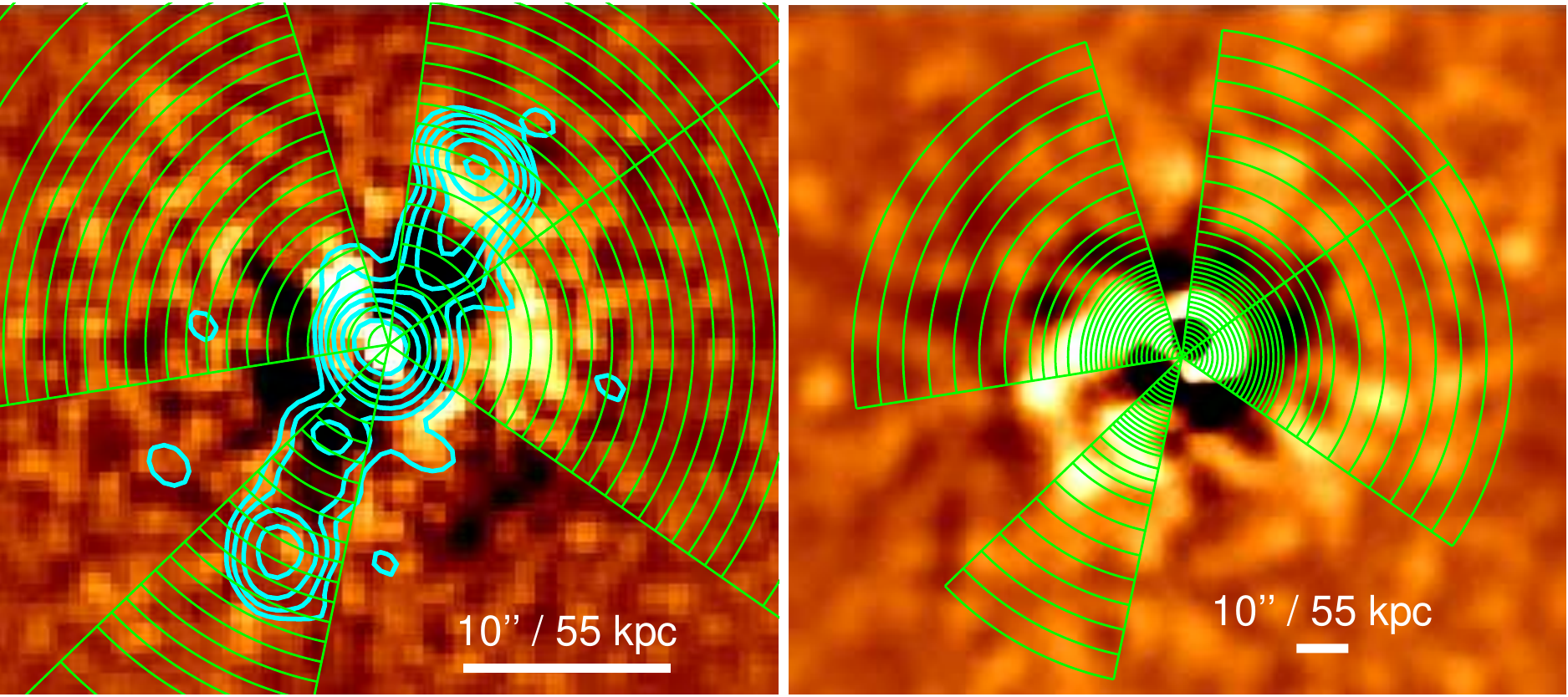}
\caption{\label{fig:resid2}\textit{Upper panels:} \chandra\ 0.3-3~keV
  (\textit{upper left}) and \xmm\ 0.5-3~keV (\textit{upper right}) residual
  images after subtraction of the best fitting elliptical $\beta$-model and
  Gaussian smoothing. The \chandra\ smoothing scale is 12 pixels
  ($\sim$6\arcs), the \xmm\ scale in 3 pixels (13.2\arcs). The two images
  have the same angular scale and alignment, and the green cross in the
  \chandra\ image indicates the optical centre of the BCG. 1.28~GHz GMRT
  contours are overlaid in green on the centre of the \xmm\ image. The same
  spiral pattern of residuals is visible in both images. The surface
  brightness hole to the northwest of the core is also visible in the
  \chandra\ image, but is too small to be resolved in the \xmms\ image.
  \textit{Lower panels:} 0.3-3~keV \chandra\ residual maps, smoothed with
  Gaussians of width 1\arcs\ (\textit{lower left}) and 6\arcs\
  (\textit{lower right}).  Cyan contours indicate the 1.28~GHz radio
  emission. Green partial annuli indicate regions used to create the radial
  profiles shown in Fig.~\ref{fig:SBprofs}.  }
\end{figure*}

Heavier smoothing reveals residual features on larger scales.
Fig.~\ref{fig:resid2} shows the same \chandra\ 0.3-3~keV image smoothed
with a $\sim$6\arcs Gaussian. The northwest cavity, surrounded by brighter
emission, is still visible as a dark hole in the image centre. However, the
heavier smoothing brings out negative residuals running from the north
around the west side of the core and connecting to the west side of the
southwest cavity. The weak negative residuals seen in Fig.~\ref{fig:resid1}
southwest of the core and extending east of the southeast cavity are thus
likely to be part of a larger structure, rather than cavities. The positive
residual to the northeast of the core is also found to extend further,
curving around the eastern and southern side of the core.

To check the robustness of the features revealed by the surface brightness
modelling, we performed a separate fit to a 0.5-3~keV \xmm\ image
(Fig.~\ref{fig:resid2}). The northwest cavity is not resolved, as its
$\sim$4\arcs\ radius is smaller than the \xmms\ point-spread function.
However, the spiral structures seen in the \chandra\ image are clear, and
can be traced in both positive and negative residuals curving out from the
core to $\sim$300~kpc.

\begin{figure}
\includegraphics[width=\columnwidth,viewport=0 210 535 730]{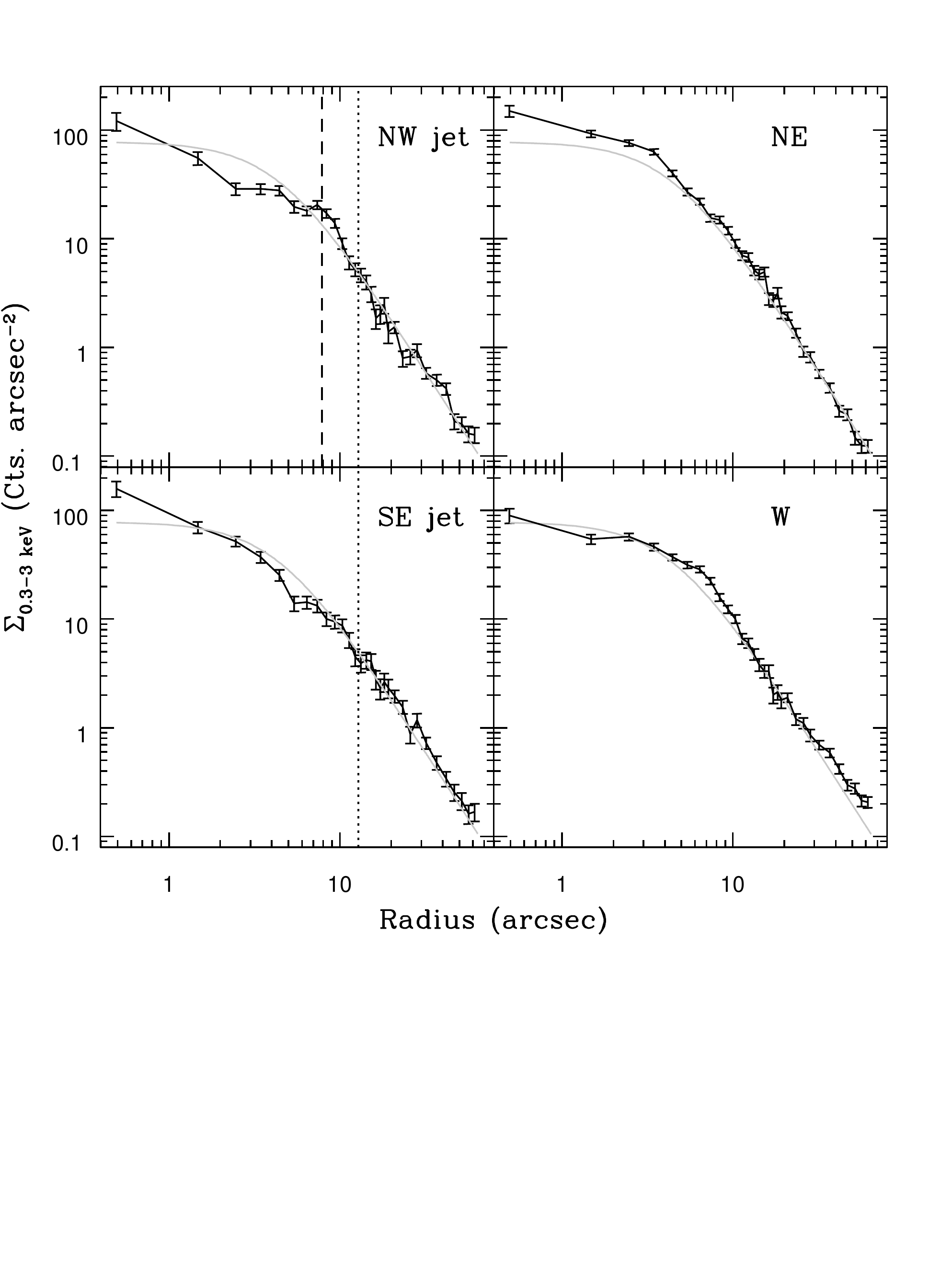}
\caption{\label{fig:SBprofs} 0.3-3~keV radial profiles derived from the
  green partial annuli marked on the Fig.~\ref{fig:resid2}. 1$\sigma$
  uncertainties are indicated with error bars, the solid grey line
  indicates the best-fitting surface brightness model to the ICM as a
  whole. Dotted lines indicate the maximum extent of the radio jets, and
  the dashed line the approximate position of the northwest filament.}
\end{figure}

Measuring the significance of the cavities and spiral structure is
difficult, since they affect the ICM as far as it can be clearly traced in
the images. We therefore compare radial profiles across the features with
the best-fitting surface brightness model. Fig.~\ref{fig:SBprofs} shows
profiles in four directions, two along the radio jets, and two roughly
perpendicular to them. Surface brightness deficits are visible at radii of
2-6\arcs\ along the northwest jet (the hole) and 5-7\arcs\ along the
southeast jet. In individual bins, the deficits are significant at the
4-5$\sigma$ level. The northwest filament is visible as a surface
brightness excess at 7-10\arcs\ radius. Neither profile shows a significant
deficit at the position of the radio lobes. A significant excess is seen at
radii $<$5\arcs\ in the northeast profile, corresponding to the residual
emission extending in this direction from the core. The western profile has
an excess at 5-10\arcs, agrees with the average surface brightness profile
at 10-20\arcs, and rises above it at larger radii. This corresponds to the
residual features observed; first a surface brightness excess on the west
side of the northwest cavity, then the region of low relative surface
brightness which forms part of the negative residual spiral in
Fig.~\ref{fig:resid2}.  These results confirm that the cavities and
associated small-scale structures are both real and significant, and
provide some support for the spiral features observed in the residual maps.

\subsection{Gas properties}
\begin{figure*}
\includegraphics[width=\textwidth, viewport=15 160 540 750]{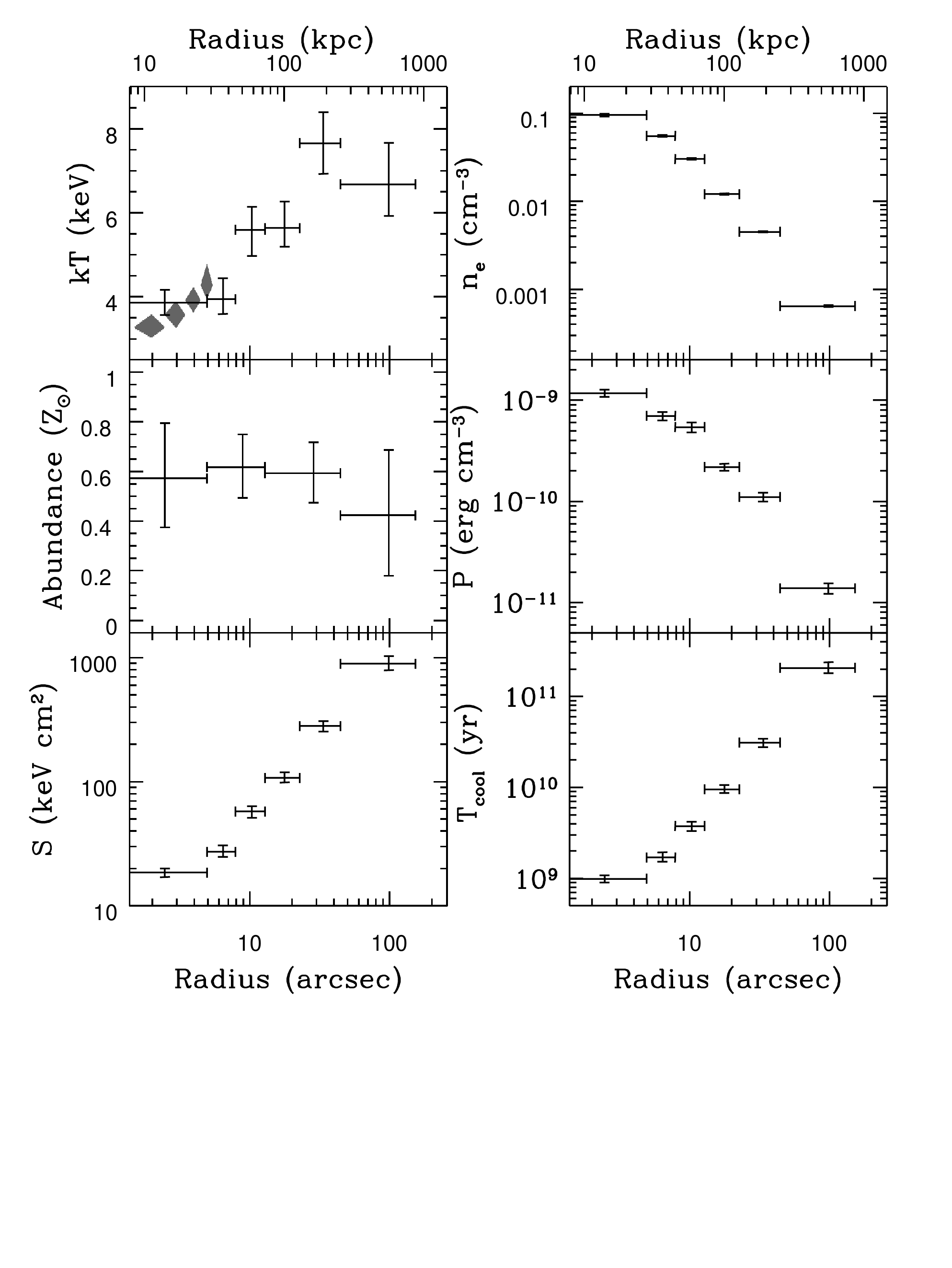}
\caption{\label{fig:Tprof} Radial profiles of gas temperature, electron
  number density (n$_e$), metal abundance, pressure (P), entropy (S) and
  isobaric cooling time (T$_{cool}$) for the cluster. Black points
  represent the deprojected fits. For comparison, projected temperatures in
  four bins at 1.5-5.5\arcs\ are marked in grey. Errorbars or regions
  indicate 1$\sigma$ uncertainties.}
\end{figure*}

To determine the physical properties of the ICM, we extracted spectra in
radial bins with widths chosen to ensure at least 3000 net source counts
per annulus. A central 1\arcs\ radius region was excluded to remove
emission from the AGN. The spectra were fitted with an absorbed APEC
thermal plasma model \citep{Smithetal01}, using the \textsc{XSpec} projct
model to account for the effects of projection. Abundances were tied
between some bins to reduce uncertainties and the typical value was found
to be $\sim$0.6\Zsol, declining outside $\sim$40\arcs\ ($\sim$220~kpc).
Fig.~\ref{fig:Tprof} shows radial profiles of the various gas
parameters derived from the fit. Our temperature, electron density and
entropy profiles are in good agreement with those of HL11. We are able to
trace the ICM to $\sim$150\arcs\ ($\sim$850~kpc) within which the cluster
has a bolometric X-ray luminosity $\sim$2$\times10^{45}$\ergps.  The radio
jets extend to $\sim$14\arcs\ (77~kpc), corresponding roughly to the outer
radius of the third annulus.  The luminosity within this radius is
$\sim$8.5$\times10^{44}$\ergps.  The peak temperature is
7.7$^{+0.7}_{-0.6}$~keV, falling to 3.9$\pm$0.3~keV in the central 5\arcs\
($\sim$28~kpc), where the cooling time is $\sim$1~Gyr.

To check for any indication of strong heating or cooling in the central
region, we also fit projected APEC models to spectra extracted from four
1\arcs-wide radial bins, starting at 1.5\arcs. We find a close agreement
between the projected and deprojected core temperatures, with no indication
of a change in temperature gradient from that observed in the deprojected
profile. In general the temperature profile shows that the heating the AGN
has delivered (either radiatively or mechanically via the jets) has been
insufficient to disrupt or destroy the strong cool core of the cluster.

We cannot accurately estimate the rate at which gas is cooling out of the
ICM, since the \xmms\ RGS data are not deep enough to allow high-resolution
spectral modelling. We can estimate mass deposition rates based on the
\chandra\ ACIS spectra, but these are very likely to be overestimated
(since the calculation neglects heating and other important physical
effects) and should be considered upper limits. Replacing the single
temperature APEC model for the central bin in the deprojection analysis
with a cooling flow model \citep[MKCFLOW,][]{MushotzkySzymkowiak88} we find
a mass deposition rate $\dot{M}$=319$^{+30}_{-34}$\Msolpyr\ within
$\sim$27~kpc. However, the fit is not a significant improvement over the
APEC model and has an unrealistic maximum temperature ($\sim$11~keV).
Alternatively, we can estimate the isobaric cooling rate:

\begin{equation}
\dot{M} \approx \frac{2\mu m_p L_{X,\rm{bol}}}{5kT},
\end{equation}

where $\mu$ is the mean molecular weight (0.593), $m_p$ is the proton mass,
$L_{X,\rm{bol}}$ is the bolometric X-ray luminosity of the central spectral
bin, and $kT$ the corresponding temperature. This gives an estimate of
$\dot{M}\sim$235\Msolpyr.

\begin{figure}
\includegraphics[width=\columnwidth]{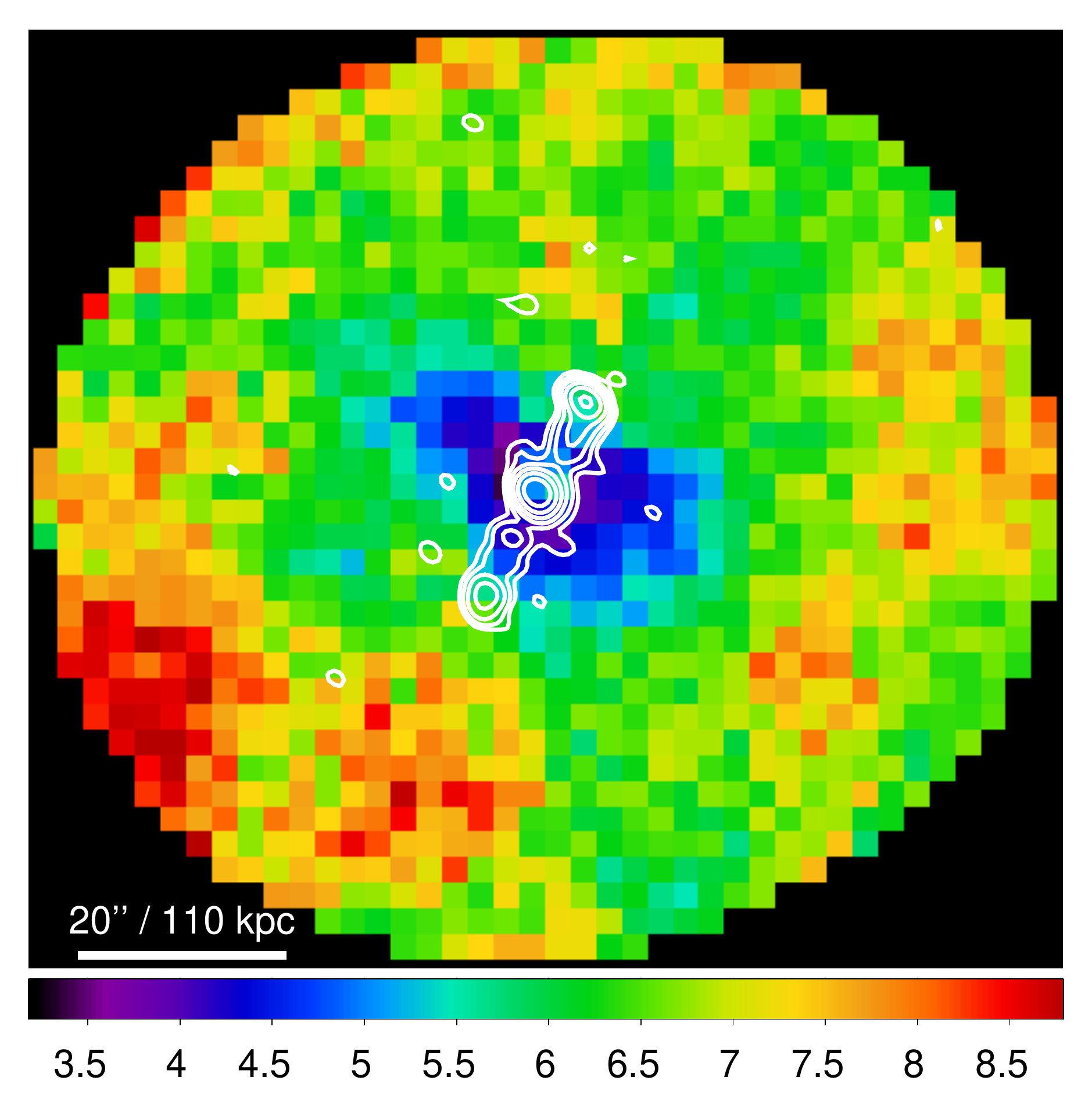}
\caption{\label{fig:tmap} Temperature map of CL09, in units of keV. Uncertainties are 7-17\%, with the central regions having smallest uncertainties. 1.28~GHz GMRT contours are overlaid.}
\end{figure}

Using the mapping technique described in \citet{OSullivanetal11a}, we
created spectral maps of the cluster core. Each 2.5$\times$2.5\arcs\ pixel
of the map represents an absorbed APEC model fitted to a spectrum extracted
from a circular region chosen to contain at least 1000 net counts. The
extraction regions had radii of 2-30\arcs\, and therefore the spectra for
neighbouring regions are generally not independent. Fig.~\ref{fig:tmap}
shows the temperature map. Uncertainties on fitted temperatures are 7-17\%.
The map shows that the coolest part of the ICM is extended along a
northeast-southwest axis, roughly perpendicular to the (projected) axis of
the radio jets, and not aligned with the major axis of the best fitting
surface brightness $\beta$-model. However, we note a good correlation
between cool temperatures and optical filaments; all the filaments are
found in regions with kT$<$5~keV, and the complex of filaments northeast of
the BCG extends along the same axis as the coolest part of the ICM.

The spectral map does not show any structures correlated with the cavities,
northwest filament, or spiral structures discussed in
section~\ref{sec:Xims}. Extracting spectra from the cavity regions shown in
Fig.~\ref{fig:resid1} and from regions chosen to cover the northwest
filament and residuals around the northern cavity, we found no significant
temperature differences beyond those expected from the map. Similarly,
placing regions to match the spiral residuals shown in
Fig.~\ref{fig:resid2}, we find only a 1.5$\sigma$ significant temperature
difference between the positive and negative residuals, the negative region
being marginally cooler (5.78$\pm$0.30~keV compared to
6.35$^{+0.22}_{-0.24}$~keV).

\subsection{Mass Profiles}
\label{sec:mass}
We determine the total gravitational mass profile of the cluster, and its
gas mass profile, using the Joint Analysis of Cluster Observations system
\citep[JACO,][]{Mahdavietal07}. This allows us to fit models jointly to the
\chandra\ and \xmm\ data, taking advantage of the improved
signal--to--noise ratio thus available. The dark matter distribution was
modelled as an NFW \citep{NavarroFW95,NavarroFW97} profile with a freely
fitted concentration parameter. For the gas profile, we used a model
consisting of three $\beta$--models, with the central component multiplied
by a power law with free slope to model the bright emission from the
central cool core. The minimum radius to which the mass profile can be
determined is limited by the point--spread function (PSF) of the \xmm\ EPIC
instruments. Since the angular size of the BCG is small ($\sim$3\arcs\
across), we cannot resolve the mass profile within the galaxy. We therefore
do not include an additional mass model to account for the stars, but
consider them to be included in the NFW model.

\begin{figure}
\includegraphics[width=\columnwidth,viewport=0 200 550 730]{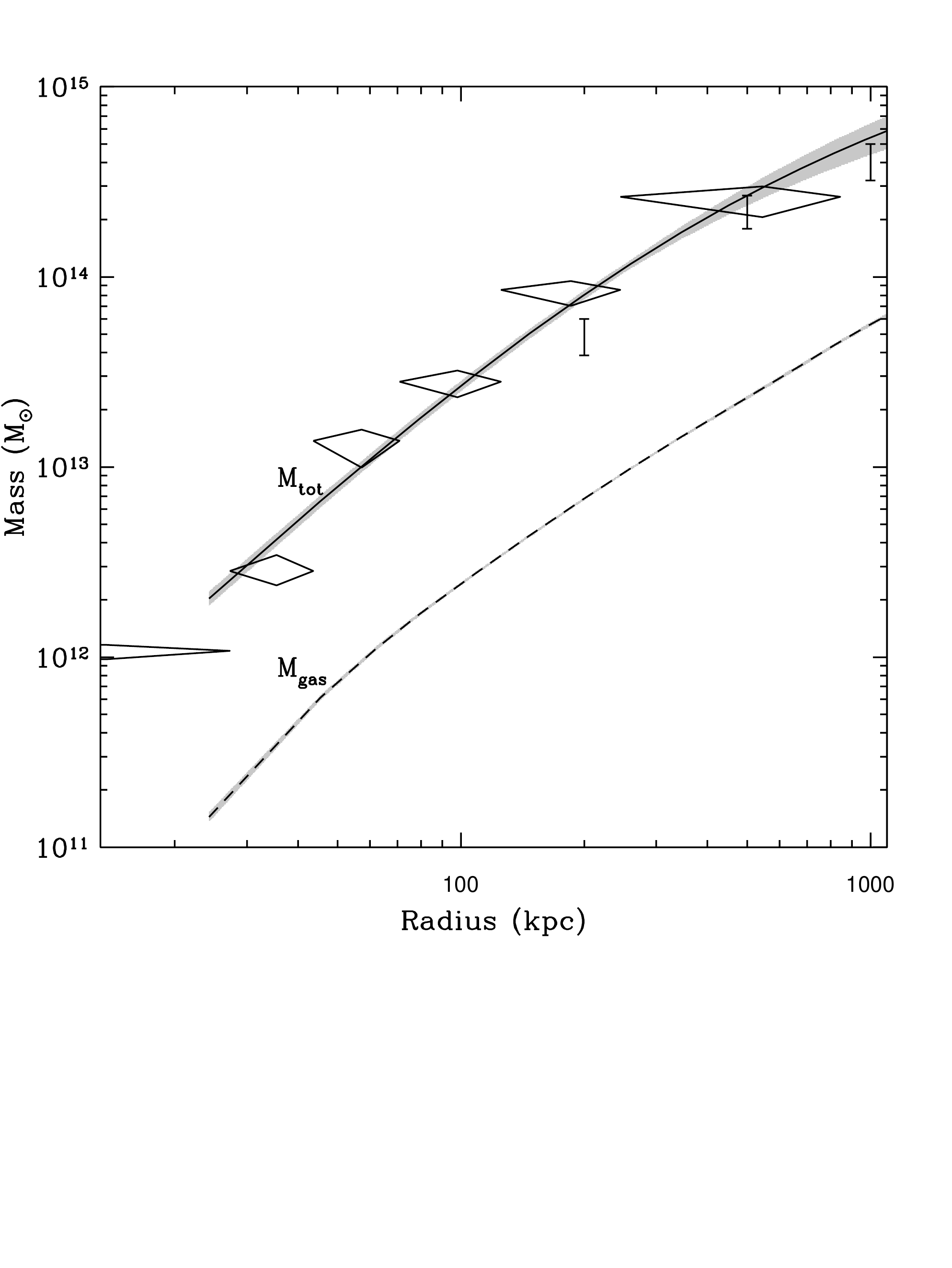}
\caption{\label{fig:mass} JACO profiles of total gravitational mass (solid line) and gas mass (dashed line) for CL09, extending to R$_{500}$. Grey regions indicate 1$\sigma$ uncertainties, though for gas mass these are similar to the width of the line. The profiles are derived from a joint fit to the \chandra\ and \xmm\ data, under the assumption of hydrostatic equilibrium. For comparison, we include mass points derived directly from our temperature and density profiles (diamonds) and from a weak lensing analysis (errorbars, Hoekstra et al. in prep.).}
\end{figure}

The results of this fitting are shown in Fig.~\ref{fig:mass}. For
comparison, we have also included mass data points derived from the
temperature and density profiles shown in Fig.~\ref{fig:Tprof}. These are
calculated under the assumption of hydrostatic equilibrium, as is the case
with the JACO modelling, but using the apparent gradients between bins. The
two estimates agree reasonably well.  From the fitted model we are able to
estimate characteristic masses and radii for CL09, as well as the enclosed
gas fraction, for different degrees of overdensity. These are shown, with
1$\sigma$ uncertainties, in table~\ref{tab:mass}. The gas fraction within
R$_{2500}$ is well within the range of gas fractions found for the sample
of massive clusters studied by \citet{Allenetal08}.

\begin{table}
  \caption{\label{tab:mass}Characteristic total masses, radii and gas fractions for the cluster. Overdensity radii are derived from the X--ray mass model, X--ray (\textit{X}) and weak lensing (\textit{wl}) masses are measured within those radii and gas fractions are derived using the X--ray mass estimates.}
  \begin{center}
    \begin{tabular}{lcccc}
      Overdensity, $\Delta$ & R$_{\Delta}$ & M$_{\Delta,X}$ & M$_{\Delta,wl}$ & f$_{gas}$\\
      & (kpc) & 10$^{14}$\Msol\ & 10$^{14}$\Msol\ & \\
      \hline
      2500 &  487$\pm$29  & 2.60$\pm$0.50 & 2.14$\pm$0.43 & 0.09$\pm$0.01 \\
      500  & 1088$\pm$97  & 5.83$\pm$1.69 & 4.12$\pm$0.98 & 0.11$\pm$0.02 \\
      200  & 1648$\pm$164 & 8.10$\pm$2.64 & - & 0.12$\pm$0.03 \\
    \end{tabular}
  \end{center}
\end{table}

We also include a comparison with weak lensing mass estimates drawn from
the Canadian Cluster Comparison Project (CCCP). This is a multiwavelength
study of a sample of $\sim 50$ X-ray luminous clusters of galaxies, which
includes CL09. To determine weak lensing masses, deep imaging data were
obtained using Megacam on the Canada-France-Hawaii Telescope.  The weak
lensing analysis is described in Hoekstra et al. (in prep).  A detailed
discussion of the analysis and the various issues that arise when
interpreting the data is provided in \citet{Hoekstra07}, which analysed an
initial sample of 20 clusters.

Fitting the measurements between 0.5-2~Mpc with a singular isothermal
sphere model, we find a best fit which corresponds to a velocity dispersion
of $\sigma$=974$^{+137}_{-160}$\kmps.  We also compute aperture masses,
which we convert to spherical masses under the assumption that the density
profile can be described by an NFW profile \citep[see][for
details]{Hoekstra07}. The resulting masses are marked on
Fig.~\ref{fig:mass}. Table~\ref{tab:mass} lists characteristic masses
measured at values of $r_\Delta$ determined from the X-ray analysis (note
we cannot determine $M_{200}$ this way). The uncertainties include the
statistical error caused by the intrinsic shapes of the sources and the
noise from distant large scale structure (e.g. Hoekstra et al. 2011).  The
lensing results are in good agreement with the X-ray mass determinations.

\section{Radio spectral analysis and physical parameters}
\label{sec:spec}

\subsection{The integrated spectrum}
\label{sec:sp}

\begin{figure}
  \centering
    \includegraphics[width=\columnwidth, bb=0 30 530 530]{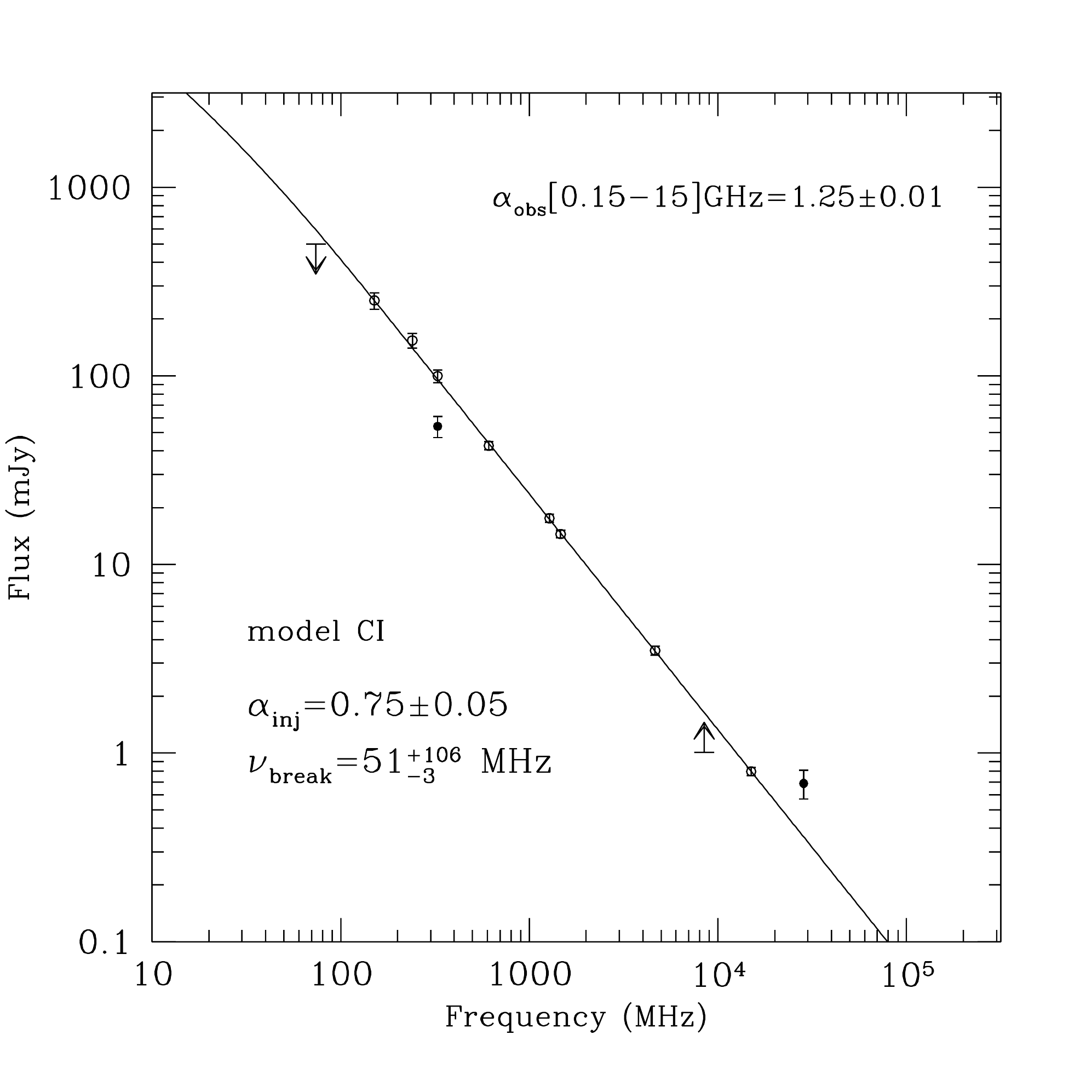}
    \caption{Integrated radio spectrum of CL09  between 74 MHz and
    4.8 GHz. The solid line is the CI fit to the data. Open points and limits indicate points included in the fit. Two points deviant points not included in the fit are marked by solid symbols.}
  \label{fig:sp}
\end{figure}

In Fig.~\ref{fig:sp} we show the integrated radio spectrum of CL09 between
74 MHz and 15 GHz, obtained using the flux densities in Table
\ref{tab:total}.  The source has a straight spectrum with steep slope,
$\alpha=1.25\pm0.01$, between 151 MHz and 15 GHz. The upper limit at 74
MHz suggests a low-frequency flattening of the spectrum, with $\alpha\le1$
below 151 MHz. We fitted the spectrum using a continuous injection (CI)
model \citep[e.g.,][]{Kardashev62}, in which electrons are continuously
injected in a region of constant magnetic field. The CI best-fit model
(solid line in Fig.~\ref{fig:sp}) provides an injection spectral index
$\alpha_{\rm inj}=0.75\pm0.05$ and a break frequency $\nu_{\rm
  break}=51^{+106}_{-3}$ MHz. Two additional data points are marked on
Fig.~\ref{fig:sp}, at 327~MHz \citep[from the WENSS
survey,][]{Rengelinketal97} and 28.5~GHz \citep{Cobleetal07}. It is unclear
why these measurements disagree with the other data points, and we
therefore do not include them in the fit.

We note that a spectral break at a frequency higher than 15 GHz is
implausible for energetic reasons. Assuming $\delta$=2$\alpha_{inj}$+1,
where $\delta$ is the slope of the injection energy spectrum of the
electrons, and $\nu_{break} \ge$ 15 GHz, the observed spectral index at
lower frequency ($\alpha$ = 1.25 $\approx \alpha_{inj}$) gives
$\delta$=3.5, which implies an extremely steep spectrum of the injected
electrons and an untenably large energy budget of the electron population,
contributed by electrons at lower energies.

\begin{figure*}
  \centering
    \includegraphics[width=16cm,angle=90]{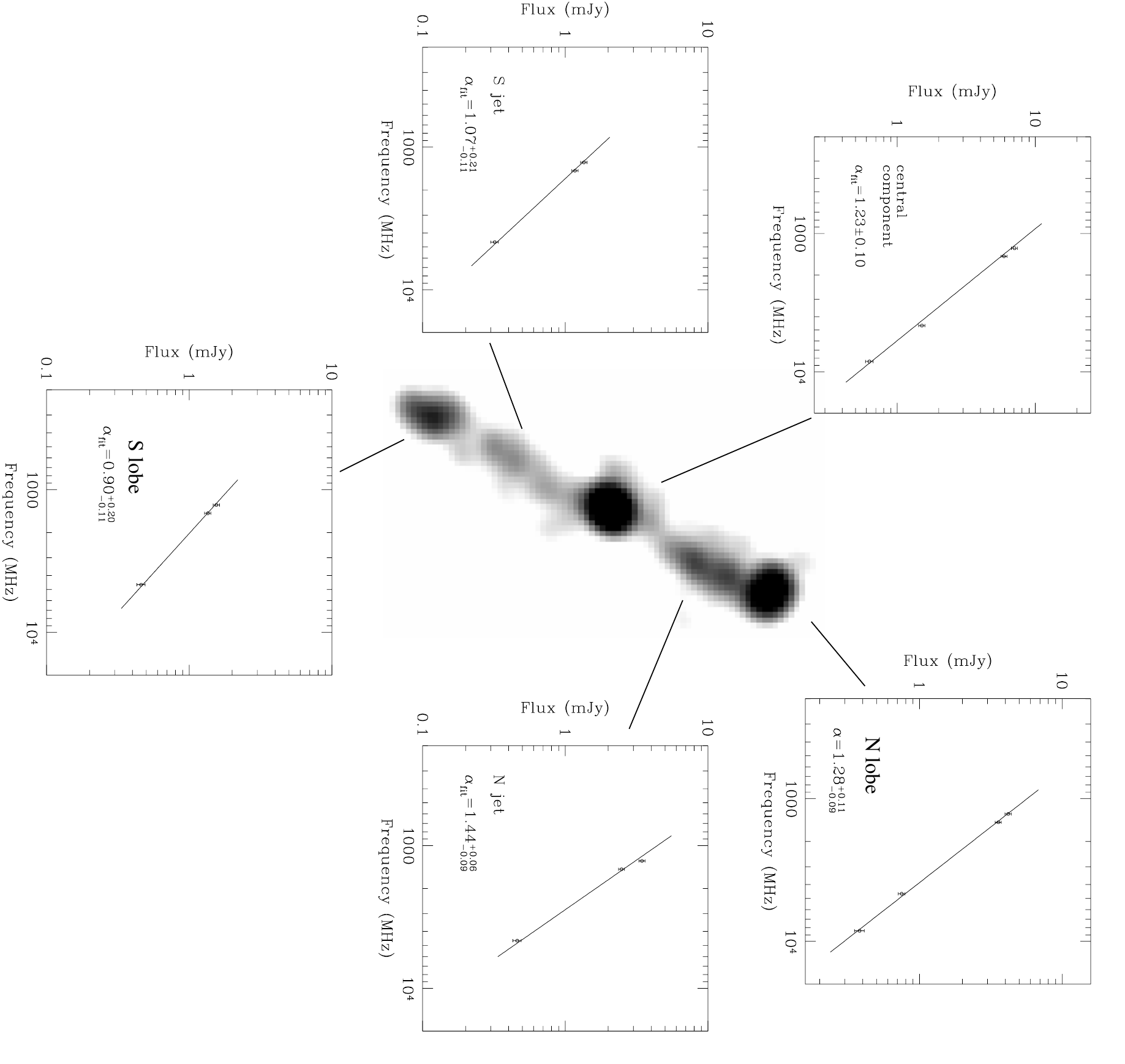}
    \caption{Integrated radio spectra of the individual components of CL09. The solid line
    in each panel is the power-law fit to the data. The image is the
    \vla\, 1.5 GHz image (same as Fig.~\ref{fig:vla}{\em a}).}
  \label{fig:spectra2}
\end{figure*}

Fig.~\ref{fig:spectra2} shows the integrated spectra of the individual
components of CL09, computed from the flux densities in
Table~\ref{tab:comp}. Given the absence of an obvious spectral curvature
and limited number of data points, we used a simple power-law model to fit
the spectrum of each component. The slopes provided by the best-fit models
are summarised in Table~\ref{tab:comp}. The spectral indices of the
individual components are all steep ($\alpha >1$), including the central
component.

\subsection{Physical parameters}\label{sec:phypar}

We calculated the physical parameters of the source using the standard
assumptions that the relativistic particle and magnetic field energy
densities are uniformly distributed over the radio source volume and in
approximate energy equipartition. We also imposed a low-energy cutoff
$\gamma_{\rm min}$ (where $\gamma$ is the electron Lorentz factor) in the
energy distribution of the radiative electrons, instead of adopting a fixed
interval of frequency \citep[typically 10 MHz-100 GHz;
e.g.,][]{Pacholczyk70}.  It has been shown that a frequency cutoff of 10
MHz may neglect the input of $\gamma \ltsim 10^3-10^4$ electrons for
typical equipartition magnetic fields of $\sim$1--10 $\mu$G
\citep{Brunettietal97,BeckKrause05}.  Since a large fraction of the
relativistic particle energy may be at lower $\gamma$, imposing a
low-energy cutoff allows to take into account the contribution from these
particles.  In our calculations, we adopt $\gamma_{\rm min}$=100, which
corresponds to a low-energy cutoff of $\sim$50 MeV.

We adopted an injection spectral index $\alpha_{\rm inj}$=0.75 provided by
the CI best--fit model (see Section~\ref{sec:sp}), and normalised the model
to the radio luminosity at 151~MHz, the lowest frequency at which the
source is detected, and thus the least affected by spectral ageing of the
electron population. Cylindrical geometry was adopted to calculate the
volume of the source. The results are summarised in Table
\ref{tab:phy_tot}, where we list the radio luminosity at 151 MHz, injection
spectral index $\alpha_{\rm inj}$, break frequency, volume, and
equipartition magnetic field $B_{\rm eq}$. The last three columns provide
estimates of the radiative age $t_{\rm rad}$, velocity growth $v_{\rm
  growth}/c$ and integrated luminosity of the source (see
Sect.~\ref{sec:age}).

\subsection{Radiative age}\label{sec:age}

Under a number of assumptions, the knowledge of the break frequency in the
spectrum of a radio source allows us to estimate the time elapsed
since the source formation \citep[e.g.,][]{MyersSpangler85}.  If radiative
losses dominate over expansion losses, the magnetic field is uniform across
the source and remain constant over the source life time, and
reacceleration processes can be neglected, the radiative age (the time
since the AGN last powered the jet, launching relativistic electrons into
it) can be obtained as

$$ t_{\rm rad} = 1590  \frac{B_{\rm eq}^{0.5}}{(B_{\rm eq}^2 + B_{\rm CMB}^2)} [(1+z) \nu_{\rm break}]^{-0.5} \,\,\,\,\,\,\textrm{Myr} $$

\noindent where $\nu_{\rm break}$ is expressed in GHz, and $B_{\rm eq}$ and
$B_{\rm CMB}$ in $\mu$G \citep{Parmaetal07}. $B_{\rm eq}$ is the
equipartition magnetic field, and $B_{\rm CMB} = 3.2(1+z)^2$ is the
equivalent magnetic field of the cosmic microwave background (CMB)
radiation, i.e., the magnetic field strength with energy density equal to
that of the CMB at the redshift $z$.

Using the break frequency in Table 5, we estimate a radiative lifetime of
$\sim$130 Myr for CL09. We also derive an estimate of the source growth
velocity $v_{\rm growth}/c$, where $c$ is the speed of light, obtained
assuming a constant velocity and a linear size of LLS=150 kpc, as measured
from the radio images in Sect.~\ref{sec:images}, i.e., $v_{\rm
  growth}=LLS/t_{\rm rad}$.  We found $v_{\rm growth} < 0.004c$
($\sim$1200\kmps). For comparison, the ICM sound speed varies from
$\sim$870-1120\kmps\ over the length of the jets.

\begin{table*}
  \caption[]{Physical parameters of CL09}
  \begin{center}
    \begin{tabular}{ccccccccc}
      \hline\noalign{\smallskip}

      log$P_{\rm 151 \, MHz}$ & $\alpha_{\rm
        inj}$ & $\nu_{break}$ & V & $B_{\rm min}$ & $t_{\rm rad}$ & $v_{\rm
        growth}$ & L$_{\rm 10~MHz-10~GHz}$\\

      (W Hz$^{-1}$) & & (MHz)  & (10$^5$ kpc$^3$) & ($\mu$G) & ($10^8$ yr) &
      ($c$) & (\ergps)\\
      \noalign{\smallskip}
      \hline\noalign{\smallskip}
      26.19  &  0.75$\pm$0.05 & 51$^{+106}_{-3}$ & 2.3  & 9.9 & 1.3 &
      0.004 & 2.1$\times$10$^{42}$ \\
      \hline{\smallskip}
    \end{tabular}
  \end{center}
  \label{tab:phy_tot}
\end{table*}

\section{Summary and Analysis of Results}
\label{sec:sum}

\subsection{The cluster}
Our X--ray observations show CL09 to be a luminous, massive cluster with a
strong cool core. The ICM is only mildly elliptical in the plane of the
sky, with no visible subclumps or near neighbours. Spectral mapping shows a
degree of ellipticity in the temperature distribution, particularly in the
core.  To some extent this ellipticity may be a product of the jet/ICM
interactions, with the inflation of cavities reducing the amount of cool
gas along the jet axis.  However, there is no indication of azimuthal
asymmetries outside the core which might invalidate our temperature profile
or mass analysis. In general, it seems fair to describe the cluster as
essentially relaxed, despite the structures detected in the ICM.

In order to look for more subtle deviations, we compared the physical
properties of CL09 to other clusters via the M--T and L--T relations. We
took the L--T relation for the REXCESS cluster survey \citep{Prattetal09},
specifically the relation between temperature and luminosity in the
0.15-1$\times$R$_{500}$ radial range. Using the value of R$_{500}$
determined in Section~\ref{sec:mass}, we extracted a \chandra\ spectrum and
fitted an absorbed APEC model. The resulting luminosity falls almost
exactly on the relation. Following the same approach, we compare with the
M--T relations of \citet{Vikhlininetal06} and \citet{Arnaudetal05}. In both
cases the cluster falls on the relation to within the 1$\sigma$
uncertainties.

The weak spiral structure found in the residual images may be an indicator of
gas motions in the ICM. Such structures are typical of sloshing motions,
where the cluster core is disturbed by the nearby passage of another
massive object \citep{AscasibarMarkevitch06,ZuHoneetal11}. Spiral
structures are observed in both surface brightness and temperature in a
number of disturbed galaxy clusters and groups
\citep[e.g.,][]{Clarkeetal04,Randalletal09b,Laganaetal10,Blantonetal11}.  We do not
observe the expected spiral structure in the temperature map of CL09, nor
do we observe the sharp paired cold fronts observed in many sloshing
systems. However, this may simply be because the \chandra\ observation has
insufficient depth to allow fine-grained temperature mapping. 

There is no clear evidence of a perturber in the X--ray data, but we can
consider what kind of system might be required to disturb the cluster.
Disturbance will be more easily caused by higher mass systems, but given
the mass of CL09, M$_{200}$=8.1$\times$10$^{14}$\Msol, it seems likely that
a group--scale system with a mass of a few 10$^{13}$\Msol\ would be
required. Such a group would be expected to have a luminosity
$>$10$^{43}$\ergps, and would thus be detected if it were still close to
CL09. As no perturber is detected, we are left with the possibility that
either the disturbing system had a very low mass (e.g., an individual
infalling galaxy or poor group), that the gas associated with the disturber
has been stripped, that it is outside the ACIS-I field of view, or that it
has already fully merged with CL09 leaving only the spiral residuals as a
record.  Simulations of cluster--cluster mergers also show spiral
structures in cases where the merger is off-axis
\citep[e.g.,][]{Pooleetal06}. The cool core and weak spiral pattern in CL09
suggest that if a merger took place, the cluster cores coalesced some time
ago and the cluster is now largely relaxed.

\subsection{The radio source}

Our analysis of the \textit{GMRT} and \textit{VLA} observations of the
large--scale radio source provide a relatively clear picture of its status.
The source is an FR~I/FR~II transition system, with relatively straight,
resolved jets and fairly small, bright lobes. The source has a steep
spectral index between 151~MHz and 15~GHz ($\alpha$=1.25$\pm$0.01), with a
probable break at $\sim$50~MHz. Separate spectral indices measured in the
jets, lobes and central component are also steep and broadly comparable,
with no clear change of spectral index with distance from the centre.  The
source is therefore probably old ($t_{rad}$=130~Myr) and no longer powered
by the AGN. There is no indication of faint, low--frequency emission on
scales larger than the jets and lobes, and therefore no indication of any
earlier radio activity.

The \textit{VLBA} observation reveals a small--scale ($\sim$200~pc) double
source, both components of which are extended. No point--like core is
detected, indicating that both components of the double source are probably
lobes, rather than being associated directly with the AGN. The flux from
the \textit{VLBA} double is comparable to the flux of the central component
of the large--scale source, with no indication of any additional emission
on other scales. The similarity in flux suggests, firstly, that the inner
double is a separate structure from the large--scale jets. There is no
indication of intermediate--scale structures between the inner double and
the base of the large--scale jets 20--30~kpc from the centre. The
well--defined, compact lobes of the inner double also support this
conclusion. Secondly, if the central component of the large--scale source
is merely the unresolved inner double, the steep spectral index measured
for the central component is also applicable to the \textit{VLBA} double,
implying that it too is old. Double sources with such small physical scales
are often assumed to be young, but it has been suggested that many such
sources may be short-lived \citep{Kunert-Bajraszewskaetal10}, and examples
where spectral index measurements show that the jets have already shut down
are not unknown \citep[e.g.,][]{Orientietal10}.  The age of the inner
double cannot be confirmed without new \textit{VLBA} observations at other
frequencies. Such observations would also determine whether the central AGN
is truly radio quiet or whether it is self--absorbed at 1.4~GHz, which is
not uncommon for radio cores.

In the plane of the sky, the axis of the VLBA inner double differs from
that of the large--scale jets by $\sim$17\degree. The similar sizes of the
two components of the inner double suggests that it is not strongly
relativistically beamed. Under the assumption that the jets are
intrinsically symmetric, the jet-to-counterjet brightness ratio can be used
to constrain the velocity $\beta c$ (where $c$ is the speed of light) and
orientation $\theta$ of the jets \citep[see][and references
therein]{Giovanninietal94}. We measure a ratio $R=1.3$ between the N and S
lobe flux densities of the double. Adopting $\alpha=1.23$, i.e., the
spectral index of the unresolved central component on arcsecond scale
(Tab.4), we estimate $\beta cos\theta$=0.03. Given the distribution of jet
velocities on the milliarcsecond scale, where usually $\beta >0.5$
\citep[e.g.,][]{Giovanninietal01}, it is very likely that the inner double
is lying almost in the plane of the sky, as $\theta$ is larger than
80\degree\ for $\beta\ge 0.2$.

\subsection{Cavity energetics and timescales}
\label{sec:cav}
From the surface brightness analysis, we have fairly strong evidence of a
cavity corresponding to the northern jet and another whose position
coincides with the southern jet but which is somewhat larger than the
current radio emission. The southern cavity lies within the negative spiral
residual, and it is therefore difficult to reliably determine its size.
Neither of these cavities is correlated with the radio lobes, only with the
jets.

\begin{table*}
  \caption{\label{tab:cav}Cavity sizes, powers and timescales}
  \begin{center}
    \begin{tabular}{lccccccccc}
      Cavity & radius & D$_{core}$ & 4$pV$ & t$_{sonic}$ & t$_{buoy}$ & t$_{refill}$ & P$_{sonic}$ & P$_{buoy}$ & P$_{refill}$ \\
      & (kpc) & (kpc) & (10$^{60}$ erg) & (Myr) & (Myr) & (Myr) & (10$^{44}$\ergps) & (10$^{44}$\ergps) & (10$^{44}$\ergps) \\
      \hline\\[-3mm]
      N & 15.5 & 22.7 & 2.2$^{+0.7}_{-0.8}$ & 20.7 & 41.9 & 122.8 & 32.8$^{+10.2}_{-12.8}$ & 16.3$^{+5.1}_{-6.3}$ & 5.5$^{+1.7}_{-2.1}$ \\ [+1mm]
      S & 22.1$\times$17.7 & 24.9 & 5.0$^{+2.4}_{-2.8}$ & 25.9 & 54.7 & 158.0 & 61.2$^{+29.6}_{-34.3}$ & 29.0$^{+14.0}_{-16.2}$ & 10.0$^{+4.8}_{-5.6}$ \\[+1mm]
      N lobe & 13.8 & 53.1 & 0.7$^{+0.2}_{-2.6}$ & 45.9 & 53.1 & 52.1 & 4.9$^{+1.5}_{-1.8}$ & 4.2$^{+1.3}_{-1.6}$ & 4.3$^{+1.3}_{-1.6}$ \\[+1mm]
      S lobe & 13.8 & 61.9 & 0.7$^{+0.2}_{-2.6}$ & 53.6 & 66.9 & 56.3 & 4.2$^{+1.3}_{-1.5}$ & 3.3$^{+1.0}_{-1.2}$ & 4.0$^{+1.2}_{-1.5}$ \\[+1mm]
    \end{tabular}
  \end{center}
\end{table*}

We might also have expected evidence of cavities coincident with the radio
lobes. Although no clear X-ray holes are seen at the positions of the
lobes, we can test whether weak surface brightness depressions might be
present. Adopting the regions shown in Fig.~\ref{fig:resid1} (circles of
radius 13.8~kpc) we compare the exposure corrected surface brightness with
equally sized regions placed to either side of the lobes, along an ellipse
defined by the best fitting surface brightness model. The southern lobe
region has a surface brightness within 1$\sigma$ of its neighbours.  The
northern lobe region is actually brighter than the neighbouring regions,
since it contains part of the northeast filament. We therefore have no
evidence of cavities coincident with the radio lobes.

We can also test whether our data are sufficiently sensitive to detect
cavities of the expected size at the position of the lobes.  We estimate
the flux deficit expected for cavities of size comparable to the radio
lobes, based on the mean density and temperature at that radius. We predict
that such cavities should produce a deficit of $\sim$30 counts (0.3-3~keV),
which is equivalent to only a 2.9$\sigma$ drop for the southern lobe. We
cannot therefore be certain that we would reliably detect cavities, if they
do exist. A comparable calculation for the northern cavity (i.e., the
X--ray hole) suggests that we should expect a decrement of $\sim$400
counts. There are no undisturbed regions at that radius which can be used
for comparison, but the observed flux is $\sim$320 counts lower than that
predicted by the best-fitting surface brightness model.

Inverse--Compton scattering of cosmic microwave background (CMB) photons by
electrons in the relativistic plasma of the radio lobes could potentially
produce additional X-ray flux in the areas where we expect to see cavities.
Under the assumption of equipartition, we estimate the energy of the
electron population of each lobe to be 5-10$\times$10$^{56}$~erg, for an
injection spectral index $\alpha_{inj}$=0.7, minimum Lorentz factor
$\gamma_{min}$=100, and an equal division of the energy in the relativistic
particle population between electrons and protons or positrons (k=1).
Following the method described in \citet{OSullivanetal10}, we estimate the
expected inverse--Compton flux from this electron population using the
relations of \citet{Erlundetal06}. The expected 0.3-3~keV flux from each
lobe is 1.3-2.2$\times$10$^{-17}$\ergpspcmsq, a factor $\sim$10$^{-4}$
lower than the flux measured in these regions. It is therefore clear that
our failure to detect cavities is not caused by inverse-Compton emission.
We note that the synchrotron self--Compton emission from the radio lobes
will be even fainter, since the energy density of radio photons in the
lobes is a factor $\sim$10$^{-3}$ lower than that of the CMB.

Since we cannot be certain that the radio lobes do not have associated
cavities, we estimated the energy required to inflate the observed
cavities, and any potential cavities at the position of the lobes. These
estimates are shown in Table~\ref{tab:cav}, as are dynamical estimates of
the timescale over which the cavities have formed and the implied
mechanical power of the radio jets.

Cavity volumes and the uncertainty on the volume are estimated following
the methods described in \citet{OSullivanetal11b}. We assume the jets to be
in the plane of the sky, and parameters from our radial spectral fits,
which account for projection effects. We define the energy of each cavity
to be 4$pV$, the enthalpy of a cavity filled with a relativistic plasma. We
estimate the time for the cavity to rise to its current position at the
speed of sound (t$_{sonic}$), the time for it to buoyantly rise to its
position \citep[t$_{buoy}$,][]{Churazovetal01}, and the time for the ICM to
refill the current cavity volume (t$_{refill}$).  For comparison, the
bolometric X-ray luminosity within the radius to which the radio jets
extend (77~kpc) is $\sim$8.5$\times10^{44}$\ergps. The mechanical energy
required to inflate the observed cavities ($pV$) is greater than this
luminosity, assuming the commonly used buoyant timescale.  Even if we use
the longer radiative age estimate ($\sim$130~Myr), the enthalpy of the
cavities (4$pV$) still exceeds the cooling luminosity. The power of the
jets is therefore more than sufficient to balance cooling in the cluster
core. If we assume that all the energy of the cavities is available to heat
the ICM, the jets can balance cooling out to at least 200~kpc (for the
refill timescale), or out to the edge of our temperature profile (for the
buoyant timescale).

Our estimates of the cavity sizes and positions differ from those of HL11.
We find the northwest cavity to be somewhat smaller (HL11 use a cavity
region which extends across the northwest X--ray filament, which we
consider likely to be the cavity rim) and the southeast cavity to be of
similar size but closer to the cluster core. These differences arise from
our different approaches to identifying cavities. HL11 examine
unsharp--masked and residual images, but select the cavity regions from the
raw image, whereas we select regions from the residual image. This
demonstrates the inherent subjectivity of cavity identification. HL11
estimate timescales and powers for the cavities which also differ from
ours, but our estimates of total jet power are in reasonable agreement,
given the uncertainties. One potentially important difference in
interpretation is that the outer edges of the cavity regions used in HL11
overlap the radio hotspots, and they therefore say that the cavities and
hotspots coincide. Our regions do not overlap the radio lobes, and both
their cavity regions and ours are poorly correlated with the radio lobes.

\begin{figure}
  \includegraphics[width=\columnwidth,viewport=0 210 540 740]{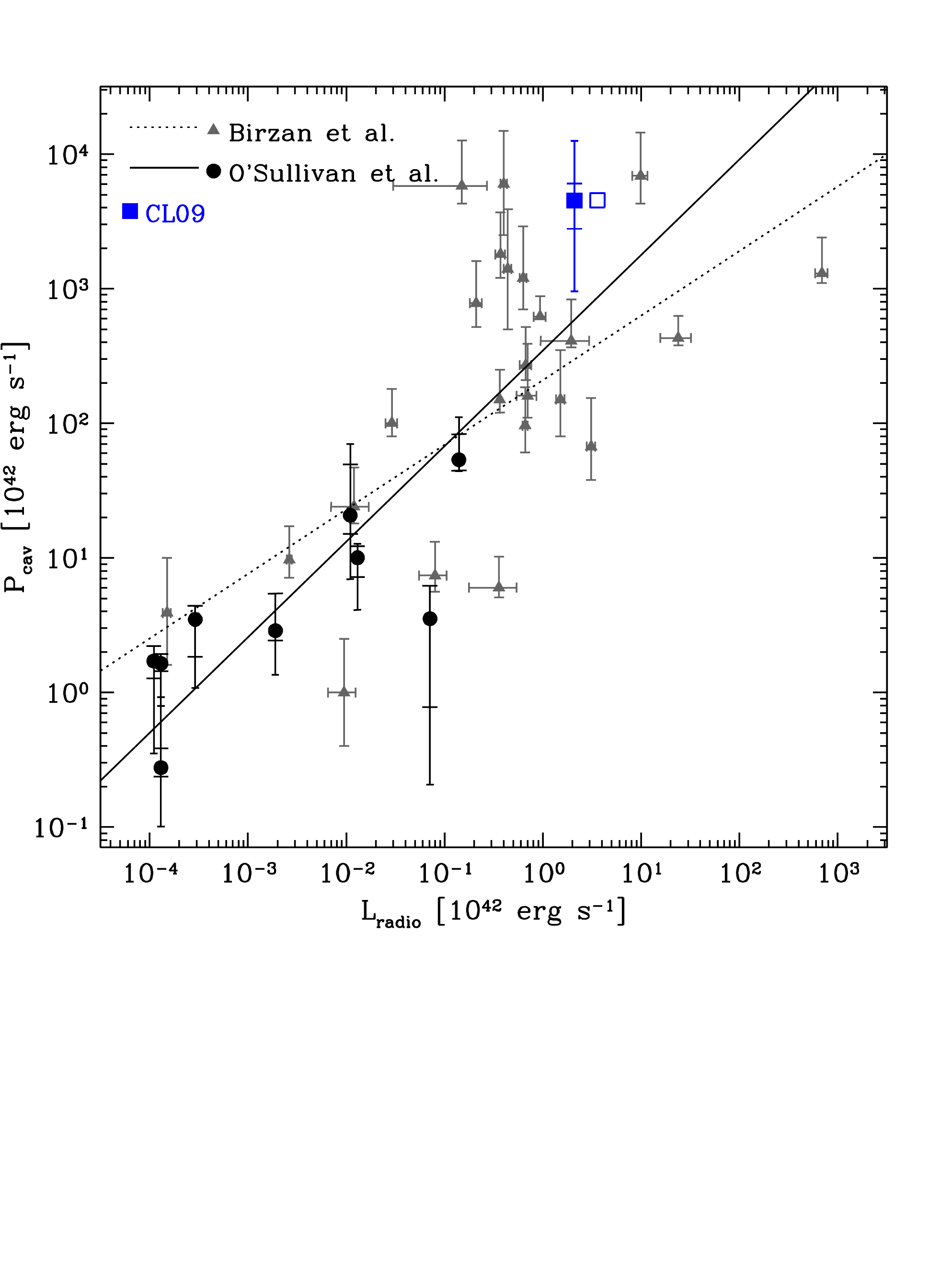}
  \caption{\label{fig:Lrad} Cavity power vs. integrated 10~MHz-10~GHz radio
    power. Systems in the sample of \citet{Birzanetal08} are marked by grey
    triangles, those in the sample of \citet{OSullivanetal11b} by black
    circles. CL09 is marked by blue squares, the filled square indicating
    the radio power observed, the open square the radio power after
    correcting for radiative ageing. 1$\sigma$ uncertainties on radio power
    and cavity power (calculated using the buoyancy timescale) are
    indicated by the error bars. For the O'Sullivan et al. groups and CL09
    additional narrow-width error bars indicate the 1$\sigma$ error range
    allowing for alternate measures of cavity age (sonic and refill
    timescales). The dotted and solid lines show the best-fitting
    relationships found by \citet{Birzanetal08} and
    \citet{OSullivanetal11b} respectively.}
\end{figure}

\begin{figure}
  \includegraphics[width=\columnwidth,viewport=0 210 540 740]{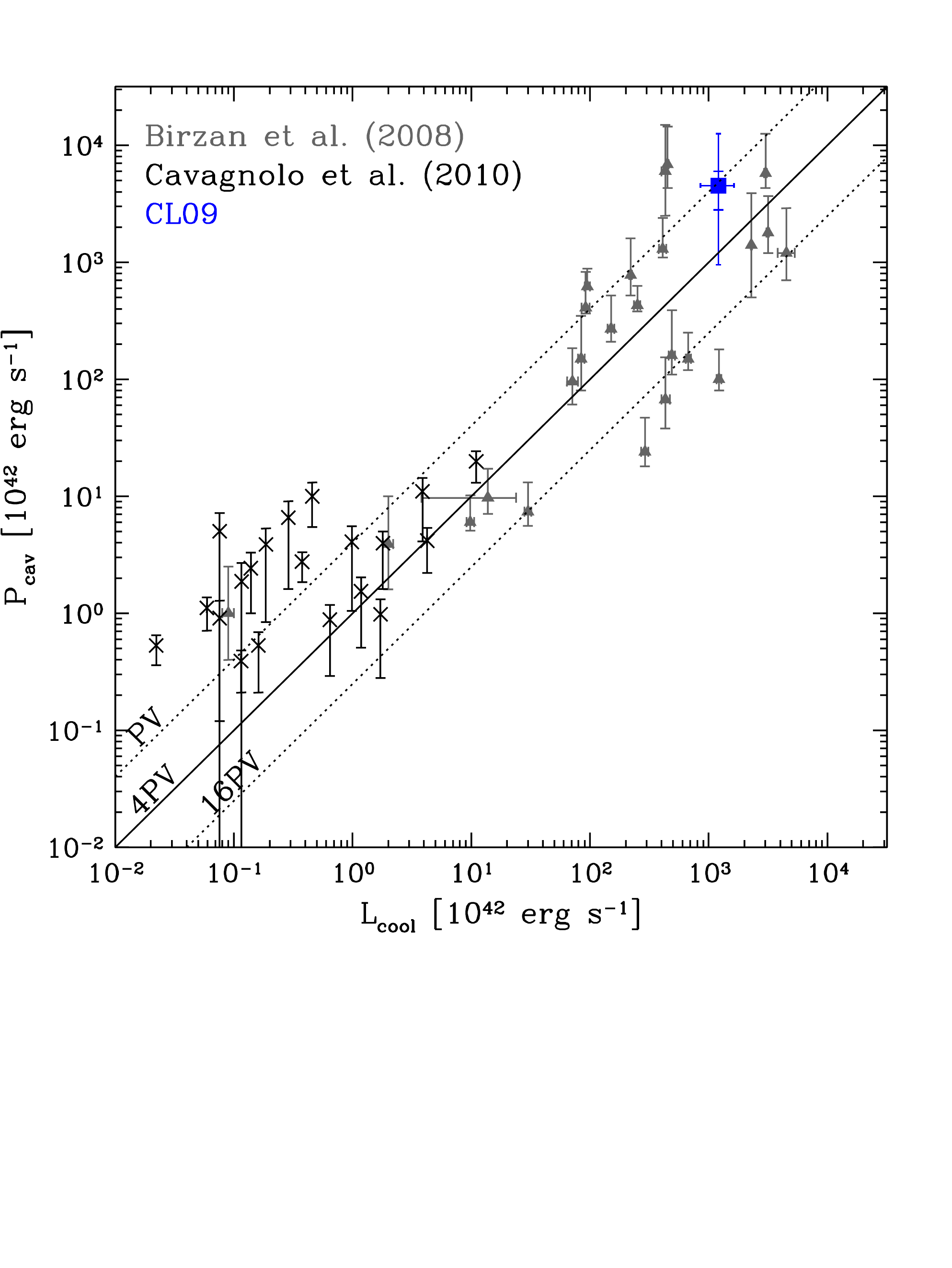}
  \caption{\label{fig:Lcool} Cavity power (4PV) against cooling luminosity
    (bolometric X-ray luminosity of the volume within which the cooling time
    $t_{cool}\leq$7.7~Gyr). The blue square indicates CL09, using the power
    estimated from the two detected cavities. Black crosses indicate the
    elliptical galaxies and groups sample of \citet{Cavagnoloetal10} and
    grey triangles the cluster sample of \citet{Birzanetal08}.  1$\sigma$
    uncertainties on cavity power, calculated using the buoyancy timescale,
    are indicated by vertical errorbars. 1$\sigma$ uncertainties on
    L$_{cool}$ are indicated by horizontal bars for CL09 and the B{\^i}rzan
    et al. sample, but in many cases the statistical uncertainties are
    smaller than the data points.  Cavagnolo et al. do not quote
    uncertainties on cooling luminosity. For CL09, additional narrow-width
    bars indicate the 1$\sigma$ uncertainties on cavity power allowing for
    alternate measures of cavity age (refill or sonic timescales) and the
    change in L$_{cool}$ between the neighbouring bins of the spectral
    profile. The solid line indicates equality between L$_{cool}$ and
    P$_{cav}$, while the dotted lines indicate equality if PV or 16PV of
    energy were required to balance cooling.}
\end{figure}

HL11 also compare CL09 to the known relations between jet power, radio
luminosity and cooling luminosity. We perform the same comparison, using
our estimates of these parameters. Fig.~\ref{fig:Lrad} shows a plot of
cavity power (for the detected cavities only) against integrated
10~MHz--10~GHz radio luminosity, which we estimate to be
2.1$\times10^{42}$\ergps. CL09 falls above the best fitting relation, but
is comparable to a number of other clusters with large jet powers. Its
offset from the relation is in part due to spectral ageing which has both
steepened the spectral index and lowered the flux we measure across all
frequencies. We can partially correct for this effect by estimating the
radio luminosity using the injection spectral index $\alpha_{inj}$=0.75
normalised to the 151~MHz flux, which (of the frequencies at which we
detect the source) is least affected by radiative ageing. We
estimate this partially corrected radio luminosity to be
3.6$\times10^{42}$\ergps. This moves the source closer to the best fitting
relation, but only by a small amount. A more important issue is the
timescale used. All data points in Fig.~\ref{fig:Lrad} use the
buoyancy timescale, but for CL09 a longer timescale may be more
appropriate, given the longer radiative age estimated from the radio
spectral analysis, and the possible link between the onset of
star--formation and shut--down of the jets (discussed in
Section~\ref{sec:history}). This would move the cluster downward, toward
the best--fitting relation. However, we note that if there are undetected
cavities coincident with the radio lobes, including them would move the
point further away from the relation.

Fig.~\ref{fig:Lcool} shows a plot of cavity power against L$_{\rm cool}$,
the luminosity of the region within which the cooling time is $<$7.7~Gyr.
CL09 has a cavity enthalpy in excess of the cooling luminosity, falling on
the $PV$ line, indicating that the mechanical power needed to inflate the
cavities is sufficient to balance cooling without any contribution from the
energy stored in the relativistic particles or magnetic field of the radio
jets.  CL09 is quite comparable to the clusters of the \citet{Birzanetal08}
sample, falling within the scatter of the cluster population. In both
Fig.~\ref{fig:Lrad} and \ref{fig:Lcool}, our results are qualitatively
similar to the findings of HL11, suggesting that the radio source in CL09
is capable of balancing radiative cooling and that it is mildly
underluminous for its jet power.

It should be noted that the jets are very likely to have expanded
supersonically for some portion of their history and would have driven
shocks into the ICM during this period. Shocks associated with radio jets
are typically weak, with Mach numbers $\la$1.5
\citep[e.g.,]{Formanetal05,McNamaraetal05}. Shocks driven by the tips of
the radio jets would have continued to move out into the ICM after the jets
shut down, becoming increasingly difficult to detect as they move into
regions of lower X--ray surface brightness. The northwest filament might
represent a shock front, or merely material pushed out subsonically by an
expanding cavity.  Unfortunately the data are too shallow to allow us to
determine whether the temperature of the northwest filament differs from
the surrounding ICM. The strength of any potential shock can be estimated
from the Rankine-Hugoniot jump conditions
\citep[e.g.,][]{LandauLifshitz59}, based on the density jump. Assuming
temperature differences have a minimal impact on surface brightness, the
density jump can be approximated as the square root of the surface
brightness jump. The surface brightness model does a rather poor job of
representing the ICM in this region, but correcting for the underlying
trend in surface brightness according to the model prediction, the
remaining surface brightness jump is only a factor of 1.08, suggesting a
shock of Mach $\sim$1.03. This suggests that the northwest filament is
likely to be the product of near-sonic expansion, rather than a supersonic
shock.

Estimating the age of the VLBA double is difficult, since its angular scale
is far smaller than that of the X-ray pressure profile. We can calculate an
approximate sonic timescale under the assumption that the central
temperature (and therefore sound speed) is similar to that observed in our
innermost projected temperature bin, $\sim$3.3~keV. This implies a sound
speed of $\sim$870\kmps\ and an timescale of order $\sim$10$^5$~yr. The
buoyant velocity depends on the gravitational mass inside the mean radius
of the lobes, which is unknown, but would exceed the sound speed unless the
enclosed mass is $\la$2.4$\times$10$^9$\Msol, and would be $\sim$2/3 of the
sound speed for 10$^9$\Msol. We have no way of recognising supersonic
expansion on these scales, so the cavities could be younger than the sonic
estimate. Conversely, we cannot confirm the presence of cavities, and if
the lobes are partially filled by entrained thermal plasma the effects of
buoyant forces could be reduced to the point where the lobes remain in the
core for long periods \citep[as in NGC~5044,][]{Davidetal09}. These
estimates of the age of the VLBA lobes should therefore be treated with
caution.

As well as the age of the radio source, we are also interested in placing
limits on the age of the current period of quasar activity. Current
simulations of AGN in groups and clusters often model the feedback as
either a bipolar outflow or a simple injection of thermal energy
\citep[e.g.,][]{Duboisetal10,McCarthyetal10,McCarthyetal11}. Both cases
produce larger--scale hot outflows which drive material into the outskirts
of the system. The current radiative luminosity of the QSO (a few
10$^{47}$\ergps) is comparable to the mechanical power of the outflows
modelled in these simulations, and the authors argue that radiative
feedback should be efficiently converted into a mechanical outflow in the
group or cluster core \citep{Duboisetal10}. If the QSO were producing a
large--scale wind, we would expect it to inflate a central cavity and/or
drive shocks out into the surrounding ICM. We might expect to detect such a
feature once it grew to a sufficient size to be resolved. Similarly, if the
QSO were directly heating the ICM, we might expect to see a central
temperature rise. We are limited by the bright central X--ray source, but
we would expect to detect features of comparable size to our innermost
temperature bin, 13.8~kpc (2.5\arcs).  For a sound speed of $\sim$870\kmps\
this suggests a timescale of $\sim$15.6~Myr. The lack of features on this
scale suggests that the AGN is either ineffective in heating the ICM or has
only been powerful enough to do so for a relatively short time. It should
also be noted that the QSO has as yet only heated the dust enshrouding it,
rather than destroying or sweeping it out of the galaxy.

\subsection{The northeast optical filament and the radio spur}
The small residual X--ray feature northeast of the core in
Fig.~\ref{fig:resid1} extends in roughly the same direction as a spur of
radio emission. Fig.~\ref{fig:spur} shows the 1.28~GHz radio and X-ray
residual contours overlaid on the \textit{HST} 622W--band image. The X-ray
and radio features appear to be correlated with the brightest of the
gaseous filaments north of the BCG.

Optically luminous filaments of ionised gas have now been observed in
numerous nearby cool core clusters, and they are known to be associated
with the coolest, densest regions of the X--ray emitting ICM
\citep[e.g.,][]{Heckmanetal89,Conseliceetal01,Fabianetal03b,Crawfordetal05,Hatchetal07,Hicksetal10,Donahueetal10,Oonketal10,McDonaldetal11}.
This association is clear in CL09, with the optical filaments falling
within the coolest part of the temperature map. This suggests that the
ionised gas is either the product of cooling from the ICM, or that it
enhances cooling through conduction of heat from the hot gas or by mixing
with the hot phase.

A correlation with radio emission is more difficult to explain.  The spur
is detected at $\sim$6$\sigma$ significance in the 1.28~GHz map, and a
similar feature on the far side of the central component is detected at
$\sim$3$\sigma$ significance. The northeast spur should be detected in the
VLA 1.5~GHz map, but is not; however a feature coincident with the
southwestern counterpart is detected. We cannot be sure whether these spurs
are real structures. A number of other small structures are seen in the
1.28~GHz map which seem likely to be noise features or marginally detected
background sources (e.g., the small blob of emission attached to the tip of
the northern lobe). We therefore consider their possible origin, but note
that deeper observations are needed to confirm or refute their existence.

Star formation is observed in some optical filaments in galaxy clusters
\citep[e.g.,][]{Johnstoneetal87,McNamaraetal93,VoitDonahue97,Crawfordetal99,ODeaetal04,Canningetal10,Oonketal11,McDonaldetal11b}, and could produce some
degree of radio emission. Measuring the 1.28~GHz flux density of the spur
is difficult, given its proximity to the bright central component, but our
best estimate is $\sim$0.25~mJy. Using the known relation between radio
flux and star formation rate \citep{Belletal03,Garnetal09}, we estimate
that an SFR of $\sim$95\Msolpyr\ would be required to produce the feature.
Star formation at this level would likely be obvious in the optical images
and would have been detected in previous spectroscopy. For comparison, \citep{Bildfelletal08} estimate an SFR or $\sim$41\Msolpyr\ for the whole BCG.

The spurs could be evidence of a third epoch of jet activity, with
a different axis to both the large--scale jets and the VLBA double. If the
spurs represent the full extent of a jet, we might expect small-scale
cavities at the position of the spurs. It is unclear whether we could
detect such small features, but the bright X--ray feature associated with
the northeast spur would certainly not be expected.

Alternatively, if the spurs were remnants of a jet which was active before
the large--scale jets but on a similar spatial scale, their radio emission
at larger radii might have faded beyond our ability to detect.  In this
case the optical filaments might be gas uplifted from the core by cavities
which have now moved out to large radii where they cannot be detected in
the \chandra\ observation. A similar correlation is seen in the Perseus
cluster, between optical filaments behind an old cavity and a ridge of
emission in the radio mini halo\citep{Fabianetal11}, but we note that there
is no evidence of any diffuse radio halo in CL09 down to the sensitivity of
our low-frequency observations. The suggestion that the spurs are a remnant
of a previous outburst is thus highly speculative and the data provide no
supporting evidence for this hypothesis beyond the presence of the spurs.

Only a deep higher resolution observation can determine whether the spurs
are real and physically correlated with the other structures in the cluster
core. We therefore conclude that while the X--ray/optical correlation is
related to cooling in the dense ICM, the reality and origin of the radio
emission is uncertain and cannot be resolved with the current data.

\section{Discussion}
\label{sec:discuss}

\subsection{History of the AGN and the cluster}
\label{sec:history}
Previous studies of CL09 have noted several features which suggest that the
orientation of the AGN, radio jets and surrounding dusty torus may have
recently changed. These include the misalignment of the radio jets and the
ionisation cone of the quasar \citep[as traced in polarized optical line
emission,][]{HinesWills93,CrawfordVanderriest96,Hinesetal99,Armusetal99,Tranetal00},
the steep high--frequency spectral index of the radio lobes, and the
flatter index and high luminosity of the radio core found by
\citet{HinesWills93}.  It was argued that the large--scale radio jets were
no longer powered by the AGN, and that some event had caused the AGN to
change its axis and launch new jets at an angle close to the line of sight,
which were as yet too small to be resolved
\citep{HinesWills93,CrawfordVanderriest96}.

\begin{figure}
  \includegraphics[width=\columnwidth]{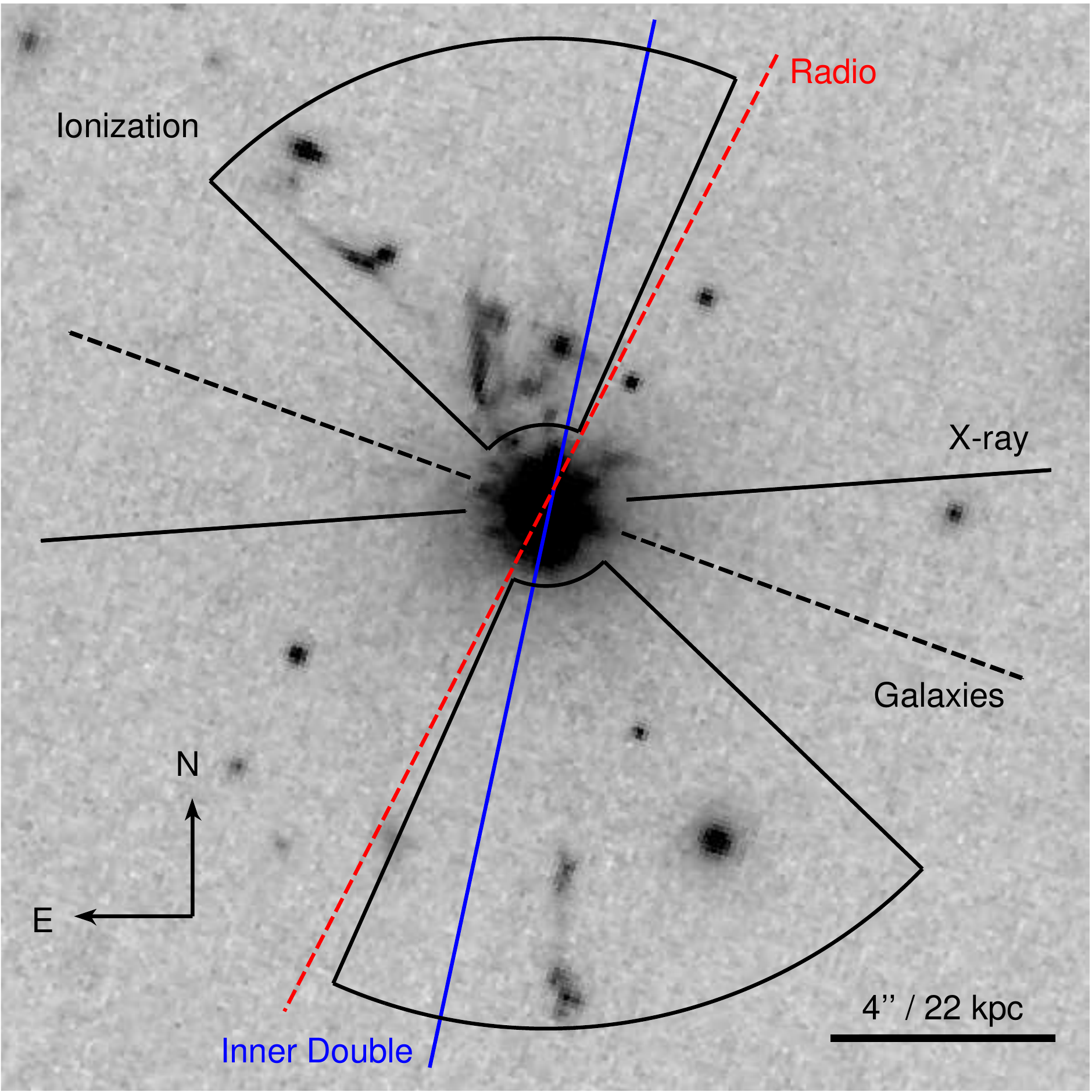}
  \caption{\label{fig:axes} \textit{HST} 622W filter image of the core of the core
    of CL09. The axes of
    the old radio jets (red dashed line), inner double radio source (blue
    solid line), major axis of the cluster X-ray emission (from our fits to
    the Chandra data, solid black lines), major axis of the cluster galaxy
    distribution \citep[black dashed lines,][]{HinesWills93} and the apparent
    opening angles of the AGN torus (solid black arcs), estimated from the
    dilution-corrected polarized optical flux \citep{Hinesetal99}, are
    indicated schematically. The position (but not the angle) of the radio
    axes has been adjusted so that they pass through the peak of optical
    surface brightness, to allow direct comparison with the torus opening
    angles. The offsets are $<$0.5\arcs.}
\end{figure}

Our results support this general picture and improve on it.
Fig.~\ref{fig:axes} shows in diagrammatic form the orientation of the
various structures associated with the AGN and BCG. As noted by
\citet{HinesWills93} the axis of the large--scale radio jets falls outside
the ionisation cones of the AGN. The VLBA inner double source has a
position angle $\sim$80\degree\ (from west), within the opening angle of
the ionisation cones and torus. However, while it was previously argued
that any new jet must be inclined close to the line of sight, the similar
sizes of the two components of the VLBA inner double suggest that it is in
fact aligned close to the plane of the sky. While the projected orientation
of the VLBA source is consistent with the ionisation cones, they do not
appear to be perfectly aligned, and an orientation close to the plane of
the sky would conflict with the inclination of $\sim$37\degree to the line
of sight found from optical polarisation measurements \citep{Hinesetal99}.
This may indicate that as well as a change in AGN orientation between the
epoch of the large--scale jets and that of the VLBA double, there has been
a further change and the inner double has shut down.

The current state of the radio jets is still open to question. Both
dynamical and synchrotron ageing arguments suggest that the large--scale
jets are old, and their spectral index confirms that they have not been
powered by the AGN for some considerable time.  The steep spectral index of
the central component of the large--scale source suggests that the inner
double is also old. In this case the lack of a resolved core in the VLBA
data and the misalignment of the inner double with respect to the
ionisation cones would be explained; the AGN would be in a radiatively
efficient accretion state with no current radio activity. However, given
the small size of the inner double ($\sim$200~pc) and short dynamical
timescale for the lobes ($\sim$10$^5$~yr), and the lack of a direct
measurement of their spectral index, it is also possible that the core is
active but self--absorbed at 1.4~GHz. Multi-frequency VLBA observations
would be needed to measure the spectral index of the inner double and
resolve this question.

\citet{HinesWills93} also noted that the radio axis is roughly aligned with
the minor axis of the distribution of galaxies within CL09 and suggested
that this correlation could be driven by the pressure gradient of the ICM.
If buoyant forces are important in the growth of the radio jets, they will
tend to align themselves with the steepest pressure gradient, in this case
along the minor axis of the cluster ICM. Fig.~\ref{fig:axes} shows that
both the large--scale radio jet and the VLBA inner double are aligned close
to the minor axis of the cluster galaxy distribution, but are somewhat less
well aligned with the ICM. The alignment may be coincidental, as the
pressure gradients on 200~pc scales are probably trivial and therefore the
alignment of the VLBA double is unlikely to be governed by buoyant forces.
It should also be noted that the coolest part of the ICM is more clearly
perpendicular to the jet axis, but this is probably caused by the
excavation of cavities by the jets, rather than the temperature
distribution affecting jet propagation.

More interestingly, the various AGN axes are poorly aligned with the BCG,
whose major axis runs almost north--south.  \citet{CrawfordVanderriest96}
note that the velocity structure of the [O\textsc{iii}] emission in the BCG
suggests rotation about a roughly east--west axis. The relatively small
rotational velocity ($\la$200\kmps) argues against the gas being located in
a simple equilibrium rotating disk given the deep potential well of the
BCG, but a rotating structure which has yet to reach equilibrium with the
galaxy seems possible.  The coincidence between the onset of the most
recent period of star formation \citep[$<$200~Myr ago][]{Pipinoetal09} and
the cessation of jet activity ($\sim$130~Myr ago), both of which could be
caused by an increase in the available cool gas supply, suggests that there
was a sudden influx of gas into the galaxy $\sim$200~Myr ago. This raises
the possibility that the BCG has recently undergone a gas--rich
merger. The presence of dust mixed with the ionised gas in the inner
filament and galaxy core \citep{Tranetal00} supports this hypothesis.
Such a merger could provide the cold gas needed to fuel the current
powerful star formation and rapidly increase the accretion rate of the AGN,
causing it to shift from a radiatively inefficient mode with large--scale
jets, to a radiatively efficient quasar mode. The axis of the AGN could
also have changed in response to the different angular momentum of the
infalling material, and might not yet have stabilised.

CL09 is a massive cluster with an extended dense ICM. A single gas--rich
galaxy falling into the cluster would likely be stripped before it could
merge with the BCG. However, a gas--rich galaxy in the core of an infalling
galaxy group or cluster could be protected by the surrounding intra-group
medium, since this would need to be stripped before galactic gas could
begin to interact with the ICM of CL09. This could potentially insulate the
galaxy until the two cluster cores merged. Simulations suggest that a
cluster--cluster merger beginning $\sim$4-5~Gyr ago could produce an
apparently relaxed cluster with weak spiral disturbances in the ICM, but
with the cluster cores merging only in the last few hundred Myr
\citep{Pooleetal06}. A merger scenario therefore seems plausible. The
distribution of galaxies could potentially provide evidence of a merger,
and it is intriguing to note that an apparent chain of galaxies extends
from northeast of the BCG across its northern side, overlapping the
northern optical filament. However, without velocity measurements from
these galaxies, we cannot be certain that this is not a mere chance
superposition.

Assuming that the star formation is taking place in a disk and follows the
standard Kennicutt-Schmidt relation \citep{Kennicutt89}, 

\begin{equation}
  \dot{M}_* = 0.017\times M_{gas} v_c/R
\end{equation}

we can use the star formation rate
\citep[$\dot{M}_*$=41\Msolpyr][]{Bildfelletal08} and available gas mass
\citep[$M_{gas}$$\sim$3.2$\times$10$^9$\Msol]{Combesetal11} to estimate the
radius of the star formation region, $R$. Taking the rotational velocity of
the [O\textsc{iii}] emission
\citep[$v_c$$\sim$200\kmps][]{CrawfordVanderriest96} to be representative,
we find that the disk is likely to have a radius $R$$\sim$270~pc
($\sim$0.05\arcs). This radius is considerably smaller than the
region of blue optical colours in the core of the BCG \citep[$\sim$20~kpc
radius,][]{Bildfelletal08}. The blue core is probably produced by a
relatively small mass of young stars formed during a more energetic period
of star formation shortly after the cold gas entered the BCG $\sim$100~Myr
ago. The ongoing star formation traced by the [O\textsc{ii}] emission is
probably fuelled by material left over from this initial burst, which has
had time to fall into the central regions of the BCG. A compact, dense,
gas--rich star forming disk around the AGN provides a natural explanation
for the heavy obscuration seen in X-ray and IR-NUV SED fitting
\citep[e.g.,][]{Vignalietal11}.

The presence of cavities, and in particular their correlation with the jets
but not the lobes, complicates our picture of nuclear activity. For the
southern jet it seems possible that the observed cavity is part of a larger
structure extending out to include the lobe, but that the \chandra\
observation is not deep enough to detect its outer parts. However, if the
northern cavity was formed by the jet, the rim of material around the
cavity must have been formed at the same time, with the northern jet first
inflating the cavity and then pushing through its outer rim to form a lobe
outside. Again, the northern lobe would in this case have an associated
cavity, which we fail to detect because the observation is too short and
the lobe is partly obscured by the rim of the inner cavity. This formation
of multiple cavities might be explained by variation of jet power during
the outburst, or by pre-existing structures in the ICM.

An alternative scenario is that the cavities were formed by a separate AGN
outburst, and are unrelated to the currently visible radio emission. If the
cavities were formed first, by some earlier AGN outburst, the weak
correlation between radio and X-ray structure would be explained, since the
jets need not pass through the cavities. We would require only that the jet
axis changes with time \citep[][and references therein]{Babuletal12}, or
that motions within the ICM have moved the cavities away from their
original axis \citep{Morsonyetal10}. The opposite case, in which the jets
formed first and the cavities through a second outburst, would require the
radio emission from this second jet to be disguised. It is just possible
that the apparent knots in the radio jet are in fact the lobes of some
later outburst.  However, this seems unlikely given their poor correlation
with the cavity morphology and the consistent spectral indices across the
jets and lobes.

In either case, the straightness of the radio jets and the fact that they
extend across the boundaries of the spiral X--ray residuals strongly
suggests that the position of the BCG relative to the surrounding ICM has
remained fairly constant over the last $\sim$100-150~Myr. If the BCG
position has remained relatively static, any cluster--cluster merger must
have occurred several Gyr ago, allowing time for the cluster to have
largely relaxed. It also seems likely that any infalling system which
provided the gas to fuel star--formation and the AGN was significantly less
massive than the BCG.

\subsection{The QSO as a source of feedback heating}
Energetically, it seems clear that the radio jets are capable of balancing
cooling in CL09, provided that: i) the duty cycle is relatively long, so
that there are not long periods of cooling with no jet activity, and ii)
that the energy injected by the jets can be efficiently coupled into the
ICM. These are conditions which apply to all systems in which feedback by
AGN jets has been proposed, though in CL09 the evidence of multiple epochs
of jet activity make it more likely that condition (i) is being met, at
least over timescales $\sim$50-150~Myr. The presence of cool gas and star
formation in cluster cores is thought to suggest that cooling from the hot
ICM is not been completely stopped by AGN feedback
\citep[e.g.,][]{Bildfelletal08,Pipinoetal09}. In CL09 we only require
cooling to be suppressed and, since the mechanical energy involved in
inflating the cavities ($PV$) is roughly equal to the cooling luminosity of
the cluster core, it seems that the jets are easily capable of providing
this suppression. It should also be noted that gas may have been brought
into the cluster core via a merger, rather than cooling from the ICM.

There is no evidence of heating of the hot ICM through radiative feedback.
The accretion related luminosity of the AGN \citep[a few
$\times$10$^{47}$\ergps,][]{Vignalietal11,Ruizetal10} is a factor $\sim$100
higher than our estimated jet powers, and comparable to the Eddington
luminosity for a 1.5$\times$10$^9$\Msol\ black hole, similar to those in
nearby BCGs such as M87, or $\sim$10 per cent. of the Eddington luminosity
if the black hole mass is of order 10$^{10}$\Msol. It therefore seems
likely that the AGN is close to its Eddington luminosity, and even with a
lower coupling efficiency than for jet feedback we might expect to see some
impact on the ICM. Radiative heating of the ICM might occur directly, or
via a wind or radiation-driven convection. Optical spectra of the AGN show
a blueshifted component with a velocity offset of $\sim$1300\kmps\
\citep{CrawfordVanderriest96} suggesting an outflow at least on small
scales. However, if the ICM is being heated we might expect to see
either a central temperature rise, a central cavity, or structures
associated with outflowing gas. There is no evidence of any such
structures, at least on the 2.5\arcs\ ($\sim$14~kpc) scales we can resolve
outside the central point source, and thus no evidence of heating on the
scale needed to balance cooling.

This lack of evidence could be an issue of power or of timescale.
It has been suggested that photoionization and Compton heating may only be
able to raise the temperature of gas to $\sim$1~keV \citep{Sazonovetal05}.
If radiative feedback can heat gas above this temperature, following
\citet{Siemiginowskaetal10} we can estimate the efficiency of the coupling
between the QSO radiation and the surrounding ICM. For the cool core, the
opacity of the gas to Compton scattering is $\tau=n_er_0\sigma_T$ where
$n_e\sim0.05$\pcmcu\ within a radius $r_0$=70~kpc, and $\sigma_T$ is the
Thompson scattering cross-section.  This suggests that $\sim$0.7\% of
photons will interact with gas in the core, but only the fraction of these
with energies $\ga$1~keV could heat the gas. This fraction is not well
known, but is likely to be $\la$10\%.  Taking into account these factors,
the luminosity available to heat the gas may only be a few
$\times$10$^{44}$\ergps, comparable to the cooling luminosity. The quasar
could thus be too weak to stop cooling, despite its apparent high
luminosity. Alternatively, the QSO may simply not have been in its current
state long enough to affect the cluster core. If our hypothesis of a gas
rich merger is correct, the transition to a quasar state was triggered by
an increase in accretion rate which also shut down the radio jets, and is
fuelled by the same material which fuelled star--formation.  Observational
studies of galaxies hosting moderate--luminosity supermassive black holes
have found a delay of $\sim$100~Myr between the times of greatest star
formation activity and peak AGN luminosity \citep{Schawinskietal09}. This
is supported by simulations, which suggest that the delay is caused by the
need to remove angular momentum from cold gas brought into the galaxy via
mergers \citep{Hopkins12}.  IRAS~09104+4109 is a considerably more massive
system than those considered by these simulations, but a similar delay is
certainly possible. The QSO has had at most 100-200~Myr to heat its
surroundings, and a dynamical delay in fuelling could mean that it may only
have reached its current high radiative power more recently.  However,
given the lack of any clear effect on the ICM, we must conclude that in
CL09, there is only evidence of feedback via the radio jets, and that only
this feedback mode is required to balance cooling from the hot ICM. In this
regard IRAS~09104+4109 appears similar to H1821+643 \citep{Russelletal10}
and 3C~186 \citep{Siemiginowskaetal10}, both (unobscured) quasars located
in cooling flow clusters whose radiative output appears to have had little
impact on their environment.

\section{Conclusions}
\label{sec:conc}
Using a combination of deep X--ray and multi-frequency radio observations,
we have examined the type-II QSO IRAS~09104+4109, located in the BCG of the
rich, cooling flow cluster CL~09104+4109.  Prior studies have shown that
the BCG is an HLIRG powered by a combination of AGN emission and star
formation, and hosts an old, steep spectrum FR~I-FR~II radio source which
has inflated cavities in the ICM.  Our analysis allows a more detailed
examination of the relationships between the cluster, BCG and AGN, and
provides insight into the history of the system. Our results can be
summarised as follows:

\begin{enumerate}
\item We present new GMRT observations at 240, 317, 610 and 1280~MHz, and
  reanalyse archival VLA observations at 1.5 and 4.8~GHz. We confirm the
  previous findings that the radio source has symmetrical straight jets
  ending in small lobes, with a bright central component. The radio
  spectrum of the source is an unbroken powerlaw extending down to
  $<$240~MHz, with a steep spectral index $\alpha$=1.25$\pm$0.01.
  Spectral index variations across the source are minimal, and the central
  source has a comparable index to both the northern and (within the
  uncertainties) southern lobes. The steep spectral index indicates that
  the jets are not currently active and the source is passively ageing. The
  straightness of the jets suggests that the position of the BCG relative
  to the local ICM has not significantly changed over the lifetime of the
  source.
\item Analysis of a previously unpublished archival 1.4~GHz VLBA
  observation of the central source resolves it into an extended double
  source $\sim$200~pc across. Both components are extended, suggesting that
  these are small--scale inner radio lobes, probably formed by a separate
  period of jet activity to the large--scale jets.  The dynamical timescale
  of the lobes is $\sim$10$^5$~yr.  The inner double may therefore be young
  compared to the large--scale radio jets, unless the buoyancy of the lobes
  is suppressed by mixing with entrained thermal plasma. However, while the
  spectral index of the inner double cannot be measured directly, its flux
  is identical (within uncertainties) to the flux measured for the central
  component of the large--scale source and there is no evidence of
  significant additional emission which might affect the spectral index
  measured on larger scales. If the spectral index of the central component
  is an accurate measure of the index of the inner double, the inner lobes
  are not currently powered by the AGN. No radio core is detected, but it
  is unclear whether this is because the AGN is not currently launching
  jets, or whether the core is active but self-absorbed at 1.4~GHz. The
  axis of the inner double differs from that of the large--scale jets by
  $\sim$17\degree, supporting the hypothesis the AGN axis has changed.
\item We measure deprojected radial profiles of temperature and density of
  the ICM out to $\sim$850~kpc, and determine the gravitational mass and
  gas mass profiles of the cluster. The core temperature of $\sim$4~keV and
  peak temperature of $\sim$8~keV agree well with previous estimates
  \citep{HlavacekLarrondoetal11}. The total mass profile agrees well with
  gravitational lesing measurements, and the cluster falls on both the L-T
  and M-T relations. The central cooling time within 5\arcs\
  ($\sim$27.5~kpc) is 9.9$\pm$0.9$\times10^8$~yr, and we estimate an
  isobaric cooling rate of $\la$235\Msolpyr\ within this radius. We
  map the 2-dimensional temperature structure of the cluster core, and find
  that the coolest gas (kT$\la$5~keV) has an asymmetrical distribution
  anti-correlated with the axis of the large--scale jets. The cool,
  optically-luminous gas filaments detected by \textit{HST} lie within this
  region of cool ICM temperatures.
\item We confirm the presence of X-ray cavities northwest and southeast of
  the BCG, coincident with the radio jets, at a radius $\sim$24~kpc.
  However, the precise relationship between the jets and cavities is
  unclear. The northern cavity is bounded on its west and northwest side by
  regions of excess surface brightness which may represent a shell of gas
  expelled from the cavity. The southern cavity appears larger, extending
  along the jet and to the east outside the radio contours, and lacks any
  clear shell. Cavities are not detected coincident with the radio lobes,
  but the \chandra\ exposure is probably insufficient to detect cavities of
  the expected volume. We estimate the enthalpy of the detected cavities
  ($\sim$7.7$\times$10$^{60}$~erg for both, summed) and for any undetected
  cavities at the location of the radio lobes. The northwest rim of the
  northern cavity has a relatively sharp, linear appearance, raising the
  possibility that it may be a shock.  Unfortunately the \chandra\
  observation lacks the depth to detect any temperature increase above that
  of the surrounding ICM, but based on the surface brightness increment we
  find that if the feature is a shock it is only mildly trans-sonic.
\item We find that the large--scale jets are no longer powered by the AGN,
  but are relics of activity which ceased $\sim$150~Myr ago. The radiative
  age estimate derived from the radio spectral index ($t_{rad}\sim$130~Myr)
  is in agreement with the refill timescales estimated for the cavities
  associated with the radio jets ($\sim$120-160~Myr). The radiative and
  dynamical timescales are in agreement with optical/NUV estimates of the
  likely age of the star formation in the BCG, $\sim$70-200~Myr. The
  similarity between these timescales and the presence of dust in
  theionised gas of the inner filament, and in the galaxy core, suggests
  that the galaxy experienced a sudden influx of cool gas $\sim$200~Myr
  ago, most likely as a result of a merger with a gas--rich interloper.  A
  rapid inflow of gas from such a merger could have triggered star
  formation and increased the accretion rate of the AGN, causing it to
  cease jet production and shift into a radiatively efficient accretion
  mode.
\item In both \chandra\ and \xmms\ data, subtraction of the best fitting
  elliptical surface brightness $\beta$-model leaves a spiral residual
  feature in the ICM, extending to $\sim$200~kpc. This is similar to
  features seen in other clusters and in simulations, associated with
  ``sloshing'' motion of the cluster core caused by a tidal encounter with
  another massive system, or with a previous merger. No sharp surface
  brightness or temperature features are observed, and the temperature map
  shows no correlation with the spiral residual, suggesting that the
  cluster has had time to relax. A merger with a galaxy group or cluster,
  beginning $\sim$4-6~Gyr ago with core merger occurring 0.5-1~Gyr ago,
  might explain both this spiral feature and the evidence of a recent
  gas--rich galaxy merger with the BCG.
\item We estimate the mechanical power output of the radio jets from the
  cavity enthalpies. Assuming that the jets are efficiently coupled with
  the ICM, we find that the jets are capable of balancing radiative cooling
  within the radius to which they extend (77~kpc), and probably out to at
  least 200~kpc. Cooling times at these radii are $>$9.5~Gyr, and it
  therefore seems likely that the jets are capable of maintaining the
  thermal balance of the cluster as a whole. While no direct evidence of
  gas heating by the AGN jets is observed, weak shocks of the type usually
  associated with jet sources would now be difficult to detect, given the
  age of the outburst and the relatively shallow \chandra\ data.
\item Although optical spectra of the AGN may suggest an outflow on small
  scales, we see no evidence of a large--scale radiatively driven wind of
  the sort predicted by some simulations of quasars in dense environments.
  The lack of any central cavity, temperature rise, or evidence of
  outflowing gas suggests that either the quasar is insufficiently
  luminous, that it has had insufficient time to produce detectable
  changes, or that radiative feedback is ineffective in heating the hot
  ICM. It is also notable that the quasar is still enshrouded by
  $\sim$3$\times$10$^9$\Msol\ of cool gas (inferred from CO measurements)
  and $<$2$\times$10$^7$\Msol\ of dust.  While some of this material has
  been heated and/or ionised, the quasar has not destroyed it or swept it
  out of the BCG. The estimated bolometric luminosity of the quasar is a
  few $\times10^{47}$\ergps and if our merger hypothesis is correct it
  entered its current radiatively efficient state $\sim$50-150~Myr ago.
\end{enumerate}

\medskip
\noindent{\textbf{Acknowledgments}}\\
The authors thank the anonymous referee for their comments, which have
improved the paper. We also thank D. Dallacasa, T. Clarke, G. Brunetti,
A.~J.~R.  Sanderson and D.~J. Burke for useful discussions of the data and
analysis, A. Siemiginowska for comments on the paper, and J.
Hlavacek--Larrondo for providing her estimates of cavity location. E.
O'Sullivan acknowledges support from the European Community under the Marie
Curie Research Training Network and from Chandra Award number AR1-12014X.
S. Giacintucci acknowledges the support of NASA through Einstein
Postdoctoral Fellowship PF0-110071 awarded by the {\em Chandra} X-ray
Center (CXC), which is operated by the Smithsonian Astrophysical
Observatory (SAO). A. Babul and C. Bildfell would like to acknowledge
support from NSERC through the Discovery Grant program.  A. Babul also
acknowledges the hospitality extended to him by the Astrophysics and Space
Research Group at the University of Birmingham in 2011. M. Donahue
acknowledges the support of a Chandra GO grant GO9-0143X.  \textbf{Chandra}
observation ObsID 10445 was proposed by K. Cavagnolo (now at 3B Scientific
in Atlanta, GA), MD, and G. M. Voit. MD acknowledges discussions with B.
McNamara. We thank the staff of the GMRT for their help during the
observations. GMRT is run by the National Centre for Radio Astrophysics of
the Tata Institute of fundamental Research. We would also like to thank the
Lorentz Center in Leiden for hospitality and support during July 2011
during which time the present study was planned. We acknowledge the usage
of the HyperLeda database (http://leda.univ-lyon1.fr). The National Radio
Astronomy Observatory is a facility of the National Science Foundation
operated under cooperative agreement by Associated Universities, Inc.

\bibliographystyle{mn2e}
\bibliography{../paper}

\label{lastpage}
\end{document}